\RequirePackage[l2tabu,orthodox]{nag}

\documentclass[12pt,a4paper,onecolumn,oneside,notitlepage]{article}
\usepackage[left=3.25cm,right=3.25cm,top=4cm,bottom=4cm]{geometry}
\usepackage[onehalfspacing]{setspace}

\usepackage{natbib}
\usepackage[british]{babel} 
\usepackage[T1]{fontenc}
\usepackage[utf8]{inputenc}
\usepackage{hyphenat}
\usepackage{microtype}
\usepackage{verbatim}
\usepackage{dsfont}

\usepackage{amsmath}
\usepackage{amsfonts}
\usepackage{amssymb}
\usepackage{amsthm}
\usepackage{nicefrac}
\usepackage{mathtools}
\usepackage{bbm} 

\usepackage{float}
\usepackage{makeidx}
\usepackage{tabularx}
\usepackage{graphicx}
\usepackage{epstopdf}
\usepackage{booktabs}
\usepackage{pdflscape}
\usepackage{longtable}
\usepackage{seqsplit}
\usepackage{rotating}
\usepackage{adjustbox}
\usepackage{multirow}
\usepackage{multicol}
\usepackage{dcolumn}
\usepackage[section]{placeins} 
\usepackage{makecell}

\usepackage{soul}
\usepackage{blindtext}
\usepackage{todonotes}
\usepackage{appendix}
\usepackage{titlesec}
\usepackage{enumerate}
\usepackage{thmtools}
\usepackage{tcolorbox}
\usepackage{colortbl} 
\usepackage[nopar]{lipsum}
\usepackage{textcomp}
\usepackage{xcolor}
\definecolor{darkRed}{RGB}{144,0,0}
\definecolor{darkBlue}{RGB}{0,0,144}
\definecolor{darkGreen}{RGB}{0,144,0}
\definecolor{darkgray}{rgb}{0.7, 0.7, 0.7}
\definecolor{lightgray}{rgb}{0.9, 0.9, 0.9}

\usepackage{hyperref}
\hypersetup{
  colorlinks=true,
  breaklinks=true,
  bookmarksnumbered=true,
  bookmarksopen=true,
  bookmarksdepth=2,
  citecolor=darkRed,
  linkcolor=darkRed,
  urlcolor=darkGreen,
  pdftitle={A general approach},
  pdfauthor={Nicolaj Søndergaard Mühlbach}
}

\titleformat{\section}[block]{\large\bfseries\centering}{\thesection.}{0.5em}{}
\titleformat{\subsection}[block]{\normalsize\itshape}{\thesubsection.}{0.5em}{}
\titleformat{\subsubsection}[block]{\normalsize\itshape\centering}{\thesubsubsection.}{0.5em}{}

\renewcommand{\thesection}{\Roman{section}}
\renewcommand{\thesubsection}{\Alph{subsection}}
\renewcommand{\thesubsubsection}{\Alph{subsection}.\arabic{subsubsection}}

\theoremstyle{plain}

\newtheorem{definition}{Definition}

\usepackage[linesnumbered,ruled,vlined]{algorithm2e}

\SetKwInput{KwInput}{Input}        
\SetKwInput{KwOutput}{Output}       

\graphicspath{{../../figures/}{../../figures/misc/}{../../figures/stylized_facts/}{figures/}{figures/misc/}{figures/stylized_facts/}}

\makeatletter
\providecommand*{\input@path}{}
\g@addto@macro\input@path{{../../tables/}{../../tables/misc/}{tables/}{tables/misc/}}
\makeatother

\newcommand{\textquote}[1]{``#1''}

\newcommand{\VerticalSpace}{\vspace{0in}} 
\newcommand{\VerticalSpaceFloat}{\vspace{0.166667in}} 

\usepackage[shortlabels]{enumitem} 
\setlist[description]{leftmargin=\parindent,labelindent=\parindent}

\usepackage{caption}
\usepackage{subcaption}
\newcommand{\RomNum}[1]{\uppercase\expandafter{\romannumeral #1\relax}}

\long\def\symbolfootnote[#1]#2{\begingroup\def\thefootnote{\fnsymbol{footnote}}\footnote[#1]{#2}\endgroup}

\begin{document}

\begin{titlepage}
	\pagenumbering{Alph}
	\newcommand{\mytitle}{\Large{\textbf{The Rise of Age-Friendly Jobs}}}

  \thispagestyle{empty}
  \setlength{\parindent}{0cm}
  \renewcommand{\thefootnote}{\fnsymbol{footnote}}
  
   \begin{center}
   
    \mytitle\symbolfootnote[1]
    {
    \noindent
The authors are grateful to participants at the Economics of Longevity conference, London Business School 2022 and an anonymous referee for helpful comments and suggestions, and to the Department of Economics, Massachusetts Institute of Technology (MIT), the Center for Research in Econometric Analysis of Time Series (CREATES), the Dale T. Mortensen Center, Aarhus University, the Danish Council for Independent Research (Grant 0166-00020B), the ESRC (Grant T002204), and the Hewlett Foundation for research support.
    }
		\vspace*{0.75cm}

    {\normalsize      
		{Daron Acemoglu}\symbolfootnote[1]
		{Department of Economics and Center for Shaping the Future of Work, MIT.
		Email: \href{mailto:daron@mit.edu}{daron@mit.edu}.}
		\hspace{0.75cm}
		{Nicolaj Søndergaard Mühlbach}\symbolfootnote[2]
		{Department of Economics, MIT, and CREATES.
		Email: \href{mailto:muhlbach@mit.edu}{muhlbach@mit.edu}.}\\
		\hfill
		{Andrew J. Scott}\symbolfootnote[3]
		{Department of Economics, London Business School and CEPR. Corresponding Author.
		Email: \href{mailto:ascott@london.edu}{ascott@london.edu}.}
		\hfill
		}
	 
 \vspace{0.5cm} 

    {
      \normalsize
      This version: \today
      \par
    }
		\vspace{-1cm}
    \thispagestyle{empty}
    \begin{singlespace}
      \begin{abstract}
				\noindent
In 1990, one in five U.S. workers were aged over 50 years whereas today it is one in three. One possible explanation for this is that occupations have become more accommodating to the preferences of older workers. We explore this by constructing an \textquote{age-friendliness} index for occupations. We use Natural Language Processing to measure the degree of overlap between textual descriptions of occupations and characteristics which define age-friendliness. Our index provides an approximation to rankings produced by survey participants and has predictive power for the occupational share of older workers. We find that between 1990 and 2020 around three quarters of occupations have seen their age-friendliness increase and employment in above-average age-friendly occupations has risen by 49 million. However, older workers have not benefited disproportionately from this rise, with substantial gains going to younger females and college graduates and with male non-college educated workers losing out the most. These findings point to the need to frame the rise of age-friendly jobs in the context of other labour market trends and imperfections. Purely age-based policies are insufficient given both heterogeneity amongst older workers as well as similarities between groups of older and younger workers. The latter is especially apparent in the overlapping appeal of specific occupational characteristics.
			\end{abstract}
    \end{singlespace}
  
	\end{center}
 
  \vspace{0.5cm}
  
\end{titlepage}

\clearpage
\setcounter{page}{1}
\pagenumbering{arabic}

\section{Introduction}\label{sec:intro}

In the U.S., the number of people aged over 50 years has risen from 65
million in 1990 to 118 million today and is expected to reach 155 million by
2050 \citep{UN2019}. Concurrently, 28 million of the 42 million increase in U.S.
employment between 1990 and 2020, is accounted for by those aged between 50
and 74 years. As a result, workers aged over 50 years have gone from being
one in five to one in three of the workforce.

\VerticalSpace
These trends are accounted for by some combination of
available jobs becoming more \textquote{age friendly} and older workers taking up employment in occupations previously performed by younger workers. Our purpose in this paper is to investigate the role of the former---the rising age-friendliness of available jobs. Doing so requires constructing a
quantitative measure of occupational age-friendliness, a key contribution of
this paper.

\VerticalSpace
The importance of age-friendly jobs is twofold. First, by reducing the
disutility of work, they should lead to greater involvement in the labour
market for older workers, a key policy objective in an ageing society \citep{Scott2022}. Second, by creating opportunities that are most attractive to older workers, they minimise the impact on wages and employment at other ages \citep{CardLemieux2001,KatzMurphy1992}
through the operation of comparative advantage. This is especially important
given the large size of the \textquote{baby-boomer} cohort that started passing the 50-year mark in 1996. Reflecting these trends, the current OECD policy is to \textquote{promote employability of workers throughout their working lives} by \textquote{creating a supportive age-friendly working environment} \citep{OECD2019}.

\VerticalSpace
The core idea of an \textquote{age-friendly working environment} is that older workers have distinct skills and preferences. Table \ref{Tbl:onet_characteristics_by_age_selection} uses data from O*NET and shows that on average older workers are in occupations that differ in numerous ways from younger workers.\footnote{Based on employment peaking as a proportion of the population in the U.S. for the age group 45 to 49 years, this paper looks at four broad age categories: age group 15-24 (pre-career), 25-49 (prime working age), 50-64
(pre-retirement) and 65-74 (post-retirement).} For instance, they are less
likely to be in occupations that require physical exertion (physical and
psychomotor abilities, work output, pace and scheduling); are more likely to
involve responsibility for others; and experience less harsh environmental
conditions and fewer job hazards. Conversely, they are more likely to be in
occupations that provide a sense of accomplishment (recognition) and better
overall working conditions. Table \ref{Tbl:onet_characteristics_by_age_selection}, however, reflects the nature of jobs undertaken and not necessarily those older people prefer. What is required is a way of determining what makes for an age-friendly job
independent from employment data.

\begin{table}[!t]
\caption{Selected O*NET Occupational Characteristics by Age 2020}
\begin{adjustbox}{max totalsize = {\textwidth}{0.6\textheight}, center}
		\begin{tabular}{lrrrrrrrrrrrrr}
\toprule
Age &  15-24 &  25-49 &  50-64 &  65-74 \\
\midrule
Cognitive Abilities       &  0.419 &  0.451 &  0.450 &  0.449 \\
Physical Abilities        &  0.211 &  0.175 &  0.170 &  0.156 \\
Psychomotor Abilities     &  0.251 &  0.231 &  0.229 &  0.212 \\
Sensory Abilities         &  0.285 &  0.296 &  0.296 &  0.291 \\
Work Output               &  0.365 &  0.366 &  0.365 &  0.348 \\
Communication             &  0.645 &  0.693 &  0.695 &  0.704 \\
Conflictual Contact       &  0.417 &  0.406 &  0.403 &  0.405 \\
Responsibility for Others &  0.549 &  0.564 &  0.566 &  0.545 \\
Environmental Conditions  &  0.291 &  0.283 &  0.284 &  0.261 \\
Job Hazards               &  0.170 &  0.173 &  0.171 &  0.152 \\
Pace and Scheduling       &  0.456 &  0.461 &  0.466 &  0.448 \\
Recognition               &  0.317 &  0.425 &  0.422 &  0.428 \\
Working Conditions        &  0.380 &  0.498 &  0.498 &  0.496 \\
\bottomrule
\end{tabular}
	\end{adjustbox}\VerticalSpaceFloat
 \label{Tbl:onet_characteristics_by_age_selection}
 {\footnotesize
    \textit{Notes:} This table shows the average value for various O*NET occupational characteristics based on weighted employment across occupations by age group. The characteristics are drawn from various O*NET categories i.e. abilities, work activities, work styles, work context, and work values. The measure runs from 0 to 1 for each indicator with a higher number denoting the occupational attribute is more apparent. See Appendix \ref{App:Data} for additional information on O*NET and description of occupational characteristics. 
\par}
\end{table}

\VerticalSpace
A number of recent papers provide answers to this question using a variety
of methodologies. \cite{Maestasetal2018} use state-preference experiments to
elicit workers' willingness to pay for different job attributes. They find
older workers prefer jobs with greater autonomy (as measured by the ability
to set their own schedule, work by themselves, etc.) as well as jobs that involve
more moderate (relative to heavy) physical activity or more sitting.

\VerticalSpace
Using both survey data as well as the probability of employment
at age 70 years, \cite{Hudomietetal2019} also examine the job
characteristics preferred by older workers. They find that older workers
have a strong preference for flexible working, reduced job stress, and less
demanding cognitive and physical work as well as less commuting time and the
opportunity to telecommute. The appeal of flexible working for older workers
is also found by \cite{Ameriksetal2020} using survey questions. They find
older workers are much more likely to accept a job if it offers flexibility
in scheduling (60 per cent of non-working respondents would be willing to
return to work if it offers a flexible schedule) and are prepared to accept
significant declines in wages in order to find flexible work (20 per cent
would accept a 20 per cent hourly wage reduction).

\VerticalSpace
These studies help define which characteristics are important
in making an occupation \textquote{age friendly} and reassuringly these differences are reflected in Table \ref{Tbl:onet_characteristics_by_age_selection}. Yet, they do not provide
information as to which \textit{specific} occupations workers consider age
friendly. To achieve this, we need to create an occupational Age-Friendliness Index
(AFI), which is the main contribution of the current paper.

\VerticalSpace
Once we construct this index, we validate that it captures the relevant
dimensions of age-friendliness by comparing it to a ranking obtained from a
small-scale survey and also document that at the beginning
of our sample, in 1990, older workers are disproportionately likely to be in
age-friendly jobs (as well as at the end of 2020).

\VerticalSpace
Our main empirical finding is to show a large increase in the age-friendliness of U.S. occupations between 1990 and 2020, with the average value of our index increasing by 8 per cent over this period. Using a Blinder-Oaxaca decomposition, we show this increase is driven in part by a rise in the relative share of occupations that were more age friendly in 1990 (the between component) but mostly through each occupation on average becoming more age friendly (the
within component). We estimate that around three quarters of occupations have become more age friendly and employment in above-median age-friendly jobs has increased by 49 million. 

\VerticalSpace
The most striking pattern we document is that this notable increase in age-friendly jobs has not disproportionately benefited
older workers. Many of these age-friendly jobs have been taken up by
females and college graduates, as the occupational characteristics preferred by older workers (e.g., flexibility, office work, less strenuous demands, etc.) also appeal to these groups. Amongst older workers, the group most disadvantaged by this competition from younger groups is male non-graduates. We discuss possible reasons why this pattern may have arisen, despite the apparent comparative advantage of older workers for more age-friendly jobs.

\VerticalSpace
The rest of the paper is organized as follows. Section \ref{sec:methods} discusses the construction of our AFI, while Section \ref{sec:application} provides a validation of this index by comparing it to rankings
by survey respondents and documenting its ability to predict occupations
with higher shares of older workers. Section \ref{sec:discussion} documents that U.S. occupations have become more age friendly, but also shows the puzzling feature that this has not disproportionately benefited older workers. It then discusses various explanations for why more age-friendly jobs have not been taken up by older workers. Section \ref{sec:conclusion} concludes, while additional details are provided in the appendix.

\section{Constructing an Age-Friendliness Index}\label{sec:methods}

Our aim in constructing an AFI is to derive a measure of
the attractiveness of different occupations to older workers. The results of 
\cite{Maestasetal2018,Hudomietetal2019,Ameriksetal2020} show which
occupational characteristics older workers find most appealing and so our
task is to measure the extent to which these are present in
different occupations and turn this into a ranking.

\VerticalSpace
For this ranking to properly capture the age-friendliness of occupations, it
should reflect not just the overall desirability of occupations for older
workers but also their \textit{relative} desirability compared to other ages. If occupations appeal particularly to older workers, they should be prepared to accept lower wages in return for these attractive features. It is this channel that should create greater employment for older workers while
minimising the impact on younger cohorts.

\VerticalSpace
With this in mind, constructing an AFI requires four steps:

\begin{enumerate}[label=\alph*)]

\item \label{AFIstep1} detailed textual descriptions of the nature and
attributes of occupations,

\item \label{AFIstep2} a precise definition of the characteristics that make
a job age friendly and a measure of their \textit{absolute} importance for
older workers,

\item \label{AFIstep3} a means of weighting these characteristics to reflect
their \textit{relative} importance for older workers compared to younger
ones,

\item \label{AFIstep4} a means of determining the degree of overlap between
these textual descriptions and using this to derive an AFI.
\end{enumerate}

\subsection{Constructing an Age-Friendliness Index---Intuition}

For \ref{AFIstep1}, we use the O*NET database that provides rich textual
details on 873 occupations. Each occupation comes with a detailed
description, a measure of the relevance of more than 244 job attributes
(broken down across abilities, interests, work values, work styles, skills,
knowledge, work activities, and work context) as well as the relative
importance of a total of 16,804 occupational tasks (see Appendix \ref{App:Data} for more details or \cite{Muhlbach2020}).

\VerticalSpace
For \ref{AFIstep2}, we utilise data from \cite{Maestasetal2018}
who obtain survey preferences for various job characteristics for older
workers.\footnote{We also construct an AFI using data from \cite{Hudomietetal2019}, using solely the Occ2Vec framework of \cite{Muhlbach2020} as well as construct a PCA version from all three AFIs. There is a high degree of correlation between all these constructed AFIs so we present only those based on \cite{Maestasetal2018}}
They define nine main categories of job characteristics: Schedule Flexibility, Telecommuting, Physical Job Demands, Pace of Work, Autonomy at Work, Paid Time Off, Working in Teams, Job Training, and Meaningful Work (which we define in Definition \ref{Def:amenity_schedule_flexibility}-\ref{Def:amenity_meaningful_work} in Appendix \ref{App:Definitions}). For each of these nine categories, \cite{Maestasetal2018} provide weights based on the willingness to pay for these features in terms of a lower wage. The smallest weights are given to Meaningful Work, Job Training and Telecommuting and the highest weights to the Absence of Physical Work, PTO and Schedule Flexibility. These provide us with a measure for the absolute appeal of different occupational
characteristics for older workers.

\VerticalSpace
For \ref{AFIstep3}, we also rely on the results of \cite{Maestasetal2018}
who provide willingness to pay for job characteristics across a range of
demographic groups. We use the differences in willingness to pay for workers
aged 62+ and those aged between 25 and 34 as a measure of the relative
importance of job characteristics for older workers. This ends up placing
the highest weight on avoiding taxing physical work, teamwork and flexible
scheduling with negative weights attached to meaningful work and learning
opportunities.

\VerticalSpace
For \ref{AFIstep4}, the challenge is that whilst O*NET provides numerical
and textual details for more than 17,000 items for each occupation, these
attributes do not exactly match with the nine components of age-friendliness
that we use from \cite{Maestasetal2018}. For instance, one of the
occupational characteristics in O*NET is \textquote{stamina}, defined as \textquote{the ability to exert yourself physically over long periods of time without getting winded or out of breath}. Similarly, one of the components that influences the age-friendliness of an occupation is its level of physical demands, defined in Definition \ref{Def:amenity_physical_job_demands} in Appendix \ref{App:Definitions} as \textquote{the level and duration of physical exertion generally required to perform job tasks, such as sitting, standing, carrying, walking, climbing stairs, lifting, carrying, reaching, pushing and pulling and it also includes
strength, flexibility, dexterity, vision and endurance}.

\VerticalSpace
Clearly, the concept of \textquote{stamina} in O*NET is related to the concept of \textquote{physical demands} in AFI, but the word
itself is not mentioned in our definition of physical demands. To overcome
this problem, we use the Occ2Vec NLP approach developed by \cite{Muhlbach2020} who uses numerical methods to work out the overlap in meaning between \textquote{stamina} and \textquote{physical
demands}.\footnote{For ease of interpretation, we explain Occ2Vec here using just one O*NET attribute and one aspect of age-friendliness. In practice, Occ2Vec works out the similarity for each occupation across a weighted average of the 17,048 available occupational descriptors and a weighted average of our nine characteristics of age-friendliness rather than for each individual pairwise comparison}
It does so by allocating numbers to words, essentially giving a numerical representation to sentences. This numerical representation is modeled
using a deep neural net which picks up number patterns representing the frequency and ways in which different words are found together. Similar concepts will therefore have similar numerical representation, as reflected in high dimensional vectors which intuitively represent the weights in the neural net. Commonly referred to as text embeddings these form the crucial component of AFI. When the embedding associated with an
occupation is more similar (as measured by cosine similarity) to the
embedding associated with our definition of age-friendliness, this
occupation will have a higher AFI.

\VerticalSpace
This AFI enables us to classify occupations both at a point in time
and across time periods. For example, if $AFI_{i,t}>AFI_{j,t}$, we say that
occupation $i$ is more age friendly than occupation $j$ at time $t$ and if $
AFI_{i,t+1}>AFI_{i,t}$, we say that the occupation has become more age
friendly between $t$ and $t+1$.

\subsection{Constructing an Age-Friendliness Index --- Technical Details}

More formally, we index occupations by $i$ for $i\in \left[ n%
\right]$ and O*NET provides us with occupational data containing weights
and text definitions given by $\left\{ \left( W_{j,t},T_{j,t}\right) :j\in \left[d\right],t\in \left[T\right] \right\}$, where $W_{j,t}\in \left[ 0,1\right] ^{n}$ is an $n$
-dimensional vector of occupational weights on the $j$th occupational
descriptor whose textual definition follows from $T_{j}$ at time $t$. The descriptors contain all tasks and attributes available from O*NET. The weights summarize the importance, relevance, and frequency of the descriptor in question for each occupation. We concatenate the weight vectors into a single matrix $\mathbf{W}_t=\left( W_{1,t}\cdots W_{d,t}\right) \in \left[ 0,1\right] ^{n\times d}$. Embedding the textual definition $T_{j,t}$ of the $j$th descriptor into a $p$-dimensional vector $D_{j,t}\in \mathbb{R}^{p}$, we apply the NLP algorithm Sentence-BERT \citep{Liu2019,reimers2019}. This gives us an embedding matrix $\mathbf{D}_t=\left( D_{1,t}\cdots D_{d,t}\right) \in \mathbb{R}^{d\times p}
$ at time $t$. The occupation vector, $X_{i,t}$ for $i\in \left[ n\right] $, is then given
as an occupation-specific average of all descriptor embeddings, $\mathbf{D}_t
\in \mathbb{R}^{d\times p}$, weighted by occupation weights. Using matrix notation, the construction of occupation
embeddings occurs via \eqref{Eq:embedding_mechanism}, that is 

\begin{align}\label{Eq:embedding_mechanism}
\mathbf{X}_t=\mathbf{W}_t\mathbf{D}_t
\end{align}
where $\mathbf{X}_t=\left( X_{1,t}\cdots X{}_{n,t}\right) \in \mathbb{R}^{n\times
p}$ and $X_{i,t}\in \mathbb{R}^{p}$ represents occupation $i$ as a $p$
-dimensional vector that encodes the relevant occupational information
across a range of criteria and with weights governed by the
occupation-specific importance of each descriptor at time $t$.

\VerticalSpace
Having generated occupation vectors $X_{i,t}\in \mathbb{R}^{p}$ for $i\in \left[ n\right]$ and $t\in \left[T\right]$, the second step is to define age-friendliness and embed its definition into another $p$-dimensional vector, say $D_{0}$.\footnote{Note that we do not index the definition of age-friendliness by time as no time-variation occurs in the definition.} This can then be used to establish the association between each occupation and age-friendliness as measured by the cosine similarity between $X_{i,t}$ and $D_{0}$.

\VerticalSpace
To ensure comparability of vectors, we use the same pre-trained
Sentence-BERT to embed each definition, resulting in nine $p$-dimensional vectors, which we concatenate into the matrix $\mathbf{A}_{0}\in \mathbb{R}^{9\times p}$. Combining these embeddings of job amenities into a weighted average leads to our embedding of age-friendliness. As weights, we use the age-specific preferences estimated by \cite{Maestasetal2018} expressed in both absolute and relative form as discussed above. Averaging the two sets of weights gives us a weight vector, $\mathbf{V}_{0}\in \mathbb{R}^{9}$,
which we use to construct our final embedding of age-friendliness via \eqref{Eq:age_friendliness_embedding}, i.e., 

\begin{align}\label{Eq:age_friendliness_embedding}
D_{0}=\mathbf{A}_{0}^{\prime }V_{0},
\end{align}
with $D_{0}\in \mathbb{R}^{p}$. Lastly, we compute the age-friendliness of the $i$th occupation at time $t$ via \eqref{Eq:age_friendliness}, that is
\begin{align}\label{Eq:age_friendliness}
AFI_{i,t} = X_{i,t}^{\prime }D_{0}\quad AFI_{i,t} \in \left[-1,+1\right],
\end{align}
which is the cosine similarity between the vectors $X_{i,t}$ and $D_{0}$,
because $\left\Vert X_{i,t}\right\Vert =\left\Vert D_{0}\right\Vert =1$.

\section{An Age-Friendliness Index}\label{sec:application}

In this section, we provide some basic descriptive
statistics about the age-friendliness of U.S. employment and occupations based on AF. We also perform validation exercises to confirm that our index successfully captures relevant dimensions of what are considered age-friendly occupations.

\subsection{Which Occupations Are Age Friendly?}

\begin{table}[!t]
\caption{Top and bottom age-friendly occupations}
\begin{adjustbox}{max totalsize = {\textwidth}{0.6\textheight}, center}
		\begin{tabular}{llll}
\toprule
          & Year &                                               1990 &                                               2020 \\
 & Rank &                                                    &                                                    \\
\midrule
Top 10 & 1  &              Vocational and educational counselors &                                             Guides \\
          & 2  &                                       Proofreaders &       Transportation ticket and reservation agents \\
          & 3  &           Baggage porters, bellhops and concierges &         Receptionists and other information clerks \\
          & 4  &        Supervisors of motor vehicle transportation &                 Advertising and related sales jobs \\
          & 5  &                  Production supervisors or foremen &                      Secretaries and stenographers \\
          & 6  &  Insurance adjusters, examiners, and investigators &       Human resources and labor relations managers \\
\bottomrule
\end{tabular}
	\end{adjustbox}\VerticalSpaceFloat
 \label{Tbl:listings_topbottom10_index_maestas_from1990_to2020}
 {\footnotesize
 \textit{Notes:} This table shows top and bottom ten occupations according to our AFI in 1990 (LHS) and 2020 (RHS).
 \par}
\end{table}

Table \ref{Tbl:listings_topbottom10_index_maestas_from1990_to2020} lists the
top and bottom age-friendly occupations in 1990 and 2020 according to AFI.
The rankings are intuitive but still revealing. Many of the top age-friendly occupations involve office work and limited physical exertion, such as proofreaders, insurance adjusters, financial managers, insurance adjusters, examiners and investigators, and business and promotion agents. Many of the least age-friendly jobs involve a heavy physical component \footnote{Our algorithm is dependent on the accuracy of O*NET descriptions. If there are biases in descriptions of the physical nature of jobs e.g physical words are heavily used in relation to construction based occupations but not nursing there will be a bias in our constructed AFI.}, such as concrete
and cement workers, construction laborers, painters, construction and
maintenance occupations. Some jobs that emerge as age friendly also involve
interpersonal communication and other soft skills, such as guides and
advertising and related sales jobs in 2020. Comparing the rankings over time
shows a notable amount of persistence, especially in the least age-friendly
occupations. 

\begin{table}[!t]
\caption{Selected O*NET occupational characteristics by AFI}
\begin{adjustbox}{max totalsize = {\textwidth}{0.6\textheight}, center}
		\begin{tabular}{lrrrrrrrrrrrrr}
\toprule
AFI quartiles  &  1 &  2 &  3 &  4 & Difference 4-1 \\
\midrule
Cognitive Abilities       &  0.413 &  0.424 &  0.442 &  0.468 & 0.055\\
Physical Abilities        &  0.291 &  0.269 &  0.178 &  0.110 & -0.181\\
Psychomotor Abilities     &  0.372 &  0.333 &  0.228 &  0.157 & -0.215\\
Sensory Abilities         &  0.328 &  0.326 &  0.294 &  0.273 & -0.055 \\
Work Output               &  0.447 &  0.441 &  0.363 &  0.314 & -0.133\\
Communication             &  0.548 &  0.608 &  0.684 &  0.761 & 0.213\\
Conflictual Contact       &  0.339 &  0.379 &  0.407 &  0.436 & 0.097\\
Responsibility for Others &  0.611 &  0.617 &  0.531 &  0.551 & -0.060\\
Environmental Conditions  &  0.445 &  0.392 &  0.270 &  0.206 & -0.239\\
Job Hazards               &  0.310 &  0.278 &  0.148 &  0.109 & -0.201\\
Pace and Scheduling       &  0.561 &  0.495 &  0.440 &  0.436 & -0.125\\
Recognition               &  0.276 &  0.343 &  0.388 &  0.491 & 0.215\\
Working Conditions        &  0.396 &  0.416 &  0.475 &  0.539 & 0.143\\
\bottomrule
\end{tabular}
	\end{adjustbox}\VerticalSpaceFloat
 \label{Tbl:onet_characteristics_by_maestas_selection}
 {\footnotesize
    \textit{Notes:} This table shows the average O*NET characteristic based on employment-weighted occupations in 2020 across quartiles of our AFI, where 1 denotes the lowest
quartile of age-friendly jobs and 4 the highest quartile. Definitions of occupational characteristics are in \ref{App:Data}.
\par}
\end{table}

The role of physical activities in the least age-friendly jobs is confirmed
in Table \ref{Tbl:onet_characteristics_by_maestas_selection}, which presents
the importance of different attributes by quartile of AFI. We can see that
our index captures the attributes emphasised in our definition of age-friendly jobs and age-friendly occupations are those that involve fewer
physical activities and job hazards, are less likely to be performed under
harsh environmental conditions, and less likely to require a rapid pace. They are also more likely to involve communication and conflictual contact as well more cognitive ability and better working conditions. It is also noteworthy that the range across quartiles in Table \ref{Tbl:onet_characteristics_by_maestas_selection} is greater than in Table \ref{Tbl:onet_characteristics_by_age_selection} that is based on age of workers and not the age-friendliness of jobs. Clearly, factors other than
comparative advantage around age-friendliness influence the age structure of
employment. 

\begin{figure}[!t]
\begin{adjustbox}{max totalsize = {\textwidth}{0.45\textheight}, center}
	\includegraphics[width = \textwidth]{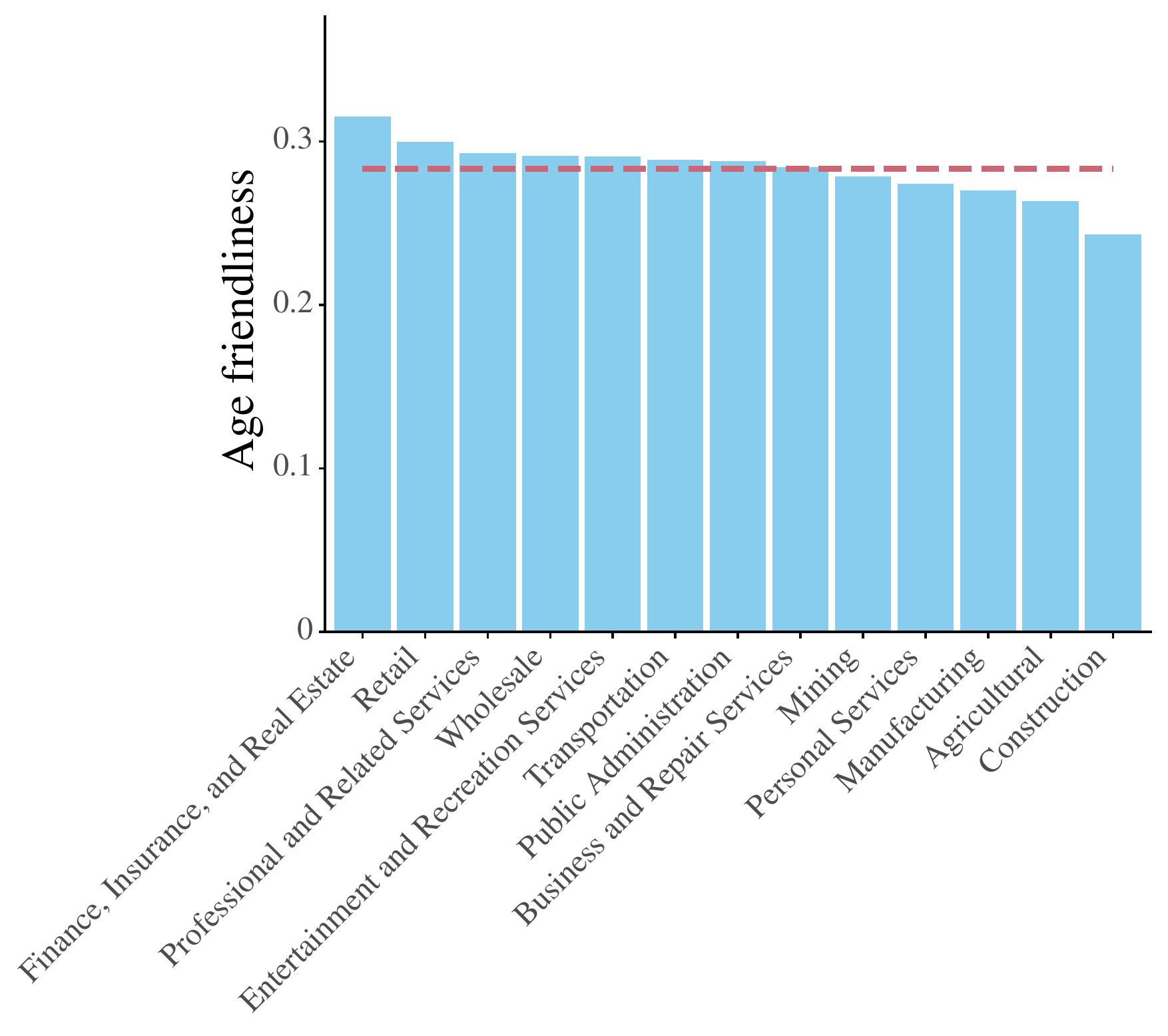}
	\end{adjustbox}
 \caption{Average Age-Friendliness Index by Industry 2020}
 \label{Fig:average_maestas_by_industry_year2020}\VerticalSpaceFloat
 {\footnotesize
 \textit{Notes:} This figure shows average age-friendliness based on employment-weighted occupations in each industry. The dashed line is the average across all sectors.
\par}
\end{figure}

\VerticalSpace
Figure \ref{Fig:average_maestas_by_industry_year2020} depicts the average
age-friendliness of occupations broken down by sector. Industries involving
harder manual work (e.g., construction and agriculture) have lower rankings
in terms of age-friendliness compared to industries with less manual work
(like finance, insurance, and real estate). As of 2020, around one in eight
workers were in an occupation in the lowest quartile of AFI and around two
in five were in a top quartile occupation. The top 10 largest occupations by
employment all score above average in terms of AFI. The largest occupation
of all is \textquote{managers and administrators},
accounting for nearly one million jobs, and it ranks in the top 10 per cent of
AFI. Males make up 69 per cent of employment in the lowest quartile AFI
occupations and 58 per cent of employment in the top
quartile AFI occupations.

\subsection{Comparison to Survey Responses}

As one part of our validation strategy, we compared the ranking according to our AFI to those of survey participants. For this purpose, we recruited a total of 210 survey participants from Amazon Mechanical Turk (MTurk), London Business
School students, and social media. Full details of recruitment and survey
composition can be found in Appendix \ref{App:Validation}.

\VerticalSpace
We ranked occupations into deciles based on AFI and then selected three
occupations from each decile. Survey participants were asked to score the
age-friendliness of ten occupations, with each survey containing an
occupation from each decile and with multiple different surveys ensuring all
30 selected occupations were ranked multiple times. To help participants
understand age-friendliness, they were provided with a definition (Definition \ref{Def:agefriendliness_survey} in Appendix \ref{App:Definitions}).\footnote{Our AFI is constructed using nine general occupational characteristics. It is the use of the relative weights based on the willingness to pay by older workers in \cite{Maestasetal2018} that makes this an age-friendliness index rather than the definition or selection of these nine characteristics. Survey participants were instead given a broad concept of what makes a job age friendly to guide them rather than the definitions of these nine occupational characteristics. This means our survey is not testing the ability of our NLP approach to successfully rank occupations based on the nine occupational characteristics but is evaluating the plausibility of our measure as reflecting general perceptions of what makes an occupation age friendly.} Participants entered their rankings into a matrix where the rows were the listed occupations and the columns an index of 1 to 10. Participants were allowed to choose \textquote{Do not know} and to give the same rank
to multiple occupations.

\begin{figure}[!t]
\begin{adjustbox}{max totalsize = {\textwidth}{0.3\textheight}, center}
		\includegraphics[width = \textwidth]{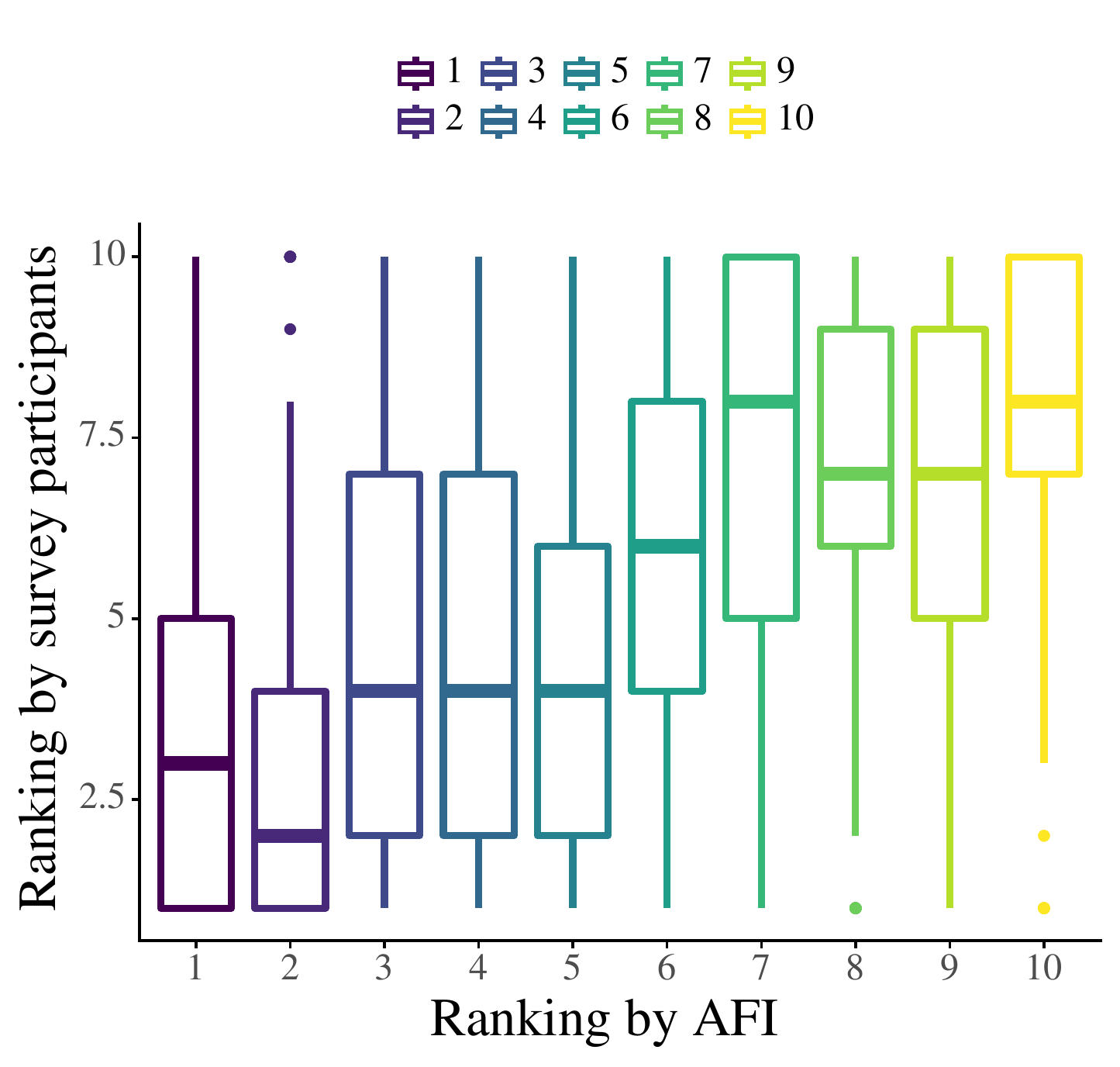}
	\end{adjustbox}	
 \caption{Proportion of deviations by age-friendliness}
 \label{Fig:Box_Whiskers}\VerticalSpaceFloat
 {\footnotesize
     \textit{Notes:} This figure shows the distribution of age-friendly rankings by survey participants compared to that based on our AFI. The horizontal bar denotes the mean ranking by survey participants, the box shows the 1st and 3rd quartile rankings and the lines the remaining distribution.
    \par}
\end{figure}

\VerticalSpace
Using the scores from participants, we construct their ranking
and compare it with that from our AFI. The Spearman rank-order correlation
coefficient shows a high degree of overlap---a
correlation coefficient of 0.829 and an associated $p$-value less than 1\textperthousand. Of course, the two rankings are not identical and the fact that there are discrepancies is revealed by mean and median absolute deviations between the two rankings of 2.2486 and 2.00, respectively. The distribution of absolute deviations is depicted in Figure \ref{Fig:Box_Whiskers}. Although there is considerable variation among the participants in terms of their rankings, the sizable correlation between the average survey rankings and those based on AFI is apparent. Moreover, looking more closely at the data, we see that absolute deviations of 0 (identical ranking) or 1 (choosing neighbouring rank) make up 45\% of the responses, and two-thirds of responses have absolute deviations of 0, 1, or 2, confirming the strong overlap between AFI and survey participants' assessments.

\subsection{Baseline Employment Patterns by AFI}

If AFI captures the comparative advantage of occupations by age---based on both the relative productivity of older workers and their relative
disutility from work---then we would expect older workers to be
disproportionately in higher AFI occupations. Table \ref{Tbl:Regression_results} confirms this by regressing the share of workers aged 50-74 in each occupation in 1990 on AFI and various other occupational characteristics. Most importantly, we control for the shares of females as well as college graduates, the average hourly wage, as well as the industry composition of
workers employed in a given occupation.\footnote{The results are similar when we separately consider the share of
workers aged 50-64 and 65-74. See Appendix \ref{App:Validation}.} In all cases, there is a strong positive correlation between AFI and the share of older workers, confirming that comparative advantage is at work in the data. Notably, the coefficient on AFI is fairly stable across columns, indicating that the relationship between AFI and the occupational share of older workers is not being driven by some other characteristics of occupations.

\VerticalSpace
That having been said, Table \ref{Tbl:Regression_results} also reveals that there are many factors other than the comparative advantage based on age-friendliness that influence the age composition of occupations. The partial $R^2$ associated with AFI is typically low and it is these deviations from comparative advantage that we discuss in the next section.

\begin{table}[!t]
\caption{AFI and Share of Older Workers in Occupations}
	\begin{adjustbox}{max totalsize = {\textwidth}{0.6\textheight}, center}
    \begin{tabular}{llllll}
\toprule
{} &        Regression \#1 &        Regression \#2 &        Regression \#3 &        Regression \#4 &        Regression \#5 \\
\midrule
AFI                                     &          $0.144^{**}$ &           $0.18^{**}$ &          $0.184^{**}$ &          $0.179^{**}$ &          $0.225^{**}$ \\
                                        &  $\left(0.073\right)$ &   $\left(0.08\right)$ &  $\left(0.081\right)$ &  $\left(0.085\right)$ &  $\left(0.091\right)$ \\
Share of females                        &                       &          $-0.026^{*}$ &          $-0.026^{*}$ &           $-0.022^{}$ &          $-0.053^{*}$ \\
                                        &                       &  $\left(0.015\right)$ &  $\left(0.015\right)$ &  $\left(0.022\right)$ &  $\left(0.028\right)$ \\
Share of graduates                      &                       &                       &           $-0.004^{}$ &           $-0.012^{}$ &            $-0.04^{}$ \\
                                        &                       &                       &  $\left(0.014\right)$ &  $\left(0.026\right)$ &  $\left(0.029\right)$ \\
Hourly wage                             &                       &                       &                       &              $0.0^{}$ &            $0.001^{}$ \\
\bottomrule
\end{tabular}
	\end{adjustbox}\VerticalSpaceFloat
\label{Tbl:Regression_results}
{\footnotesize
\textit{Notes:} This table shows results from regressing the 1990 share of employment of workers aged 50-74 in each occupation on the variables listed in column 1. Superscripts ***, **, and * indicate statistical significance based on a (two-sided) $t$-test using heteroskedasticity-robust standard errors at significance levels 1\%, 5\%, and 10\%, respectively.
\par}
\end{table}


\section{Have U.S. Jobs Become More Age Friendly?}\label{sec:discussion}

In this section, we show that even though U.S. jobs have
become more age friendly, many of these age-friendly jobs have been taken by younger workers.

\subsection{Age-Friendliness Trends}

\begin{figure}[!t]
\begin{adjustbox}{max totalsize = {\textwidth}{0.3\textheight}, center}
		\includegraphics[width = \textwidth]{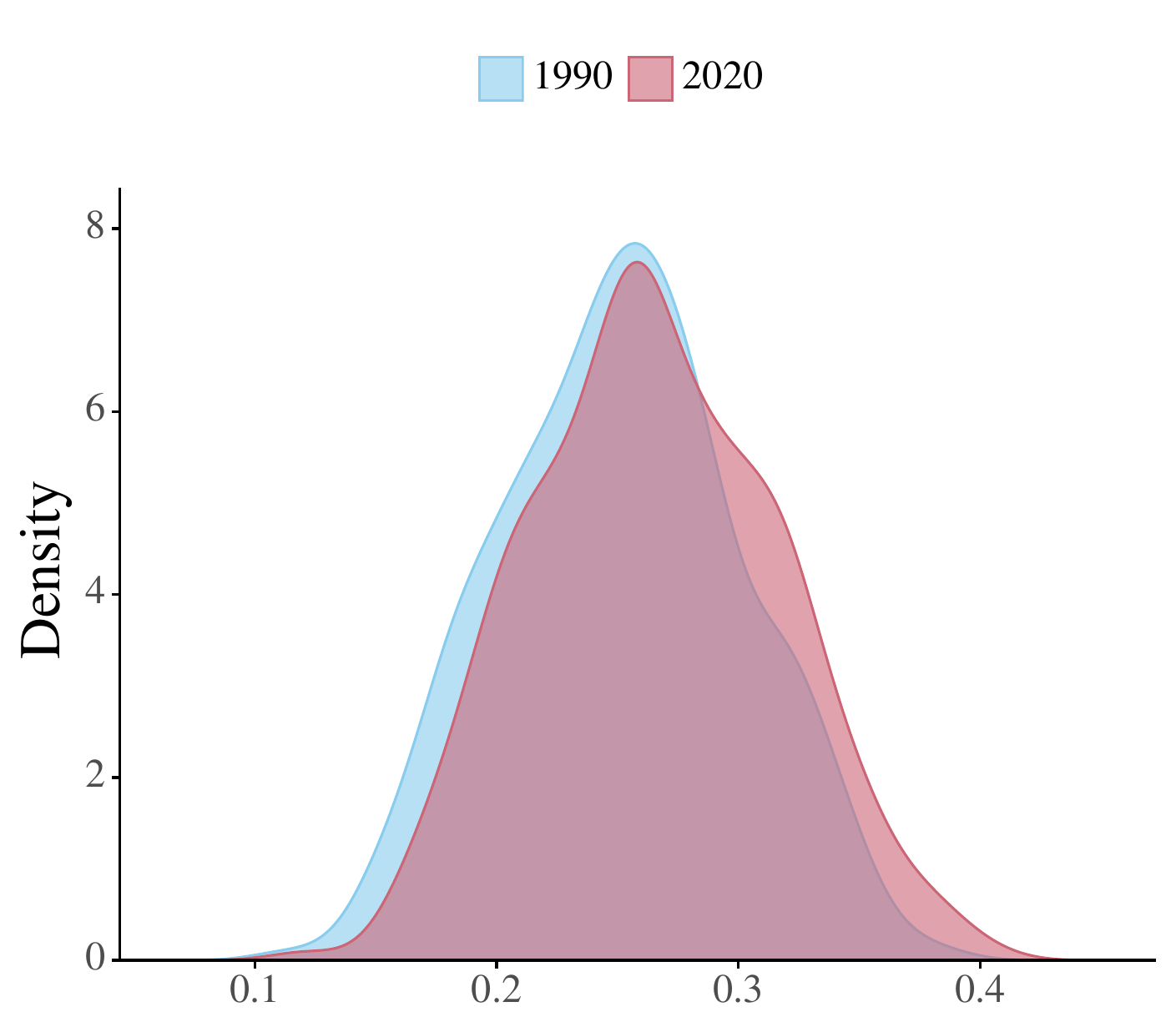}
	\end{adjustbox}
 \caption{Distribution of AFI by Occupations 1990 and 2020}
 \label{Fig:density_maestas_theoretical_year_1990_2020}\VerticalSpaceFloat
{\footnotesize
    \textit{Notes:} This figure shows the distribution of occupations (not weighted by employment) across our AFI for 1990 and 2020.
    }\par
\end{figure}

Figure \ref{Fig:density_maestas_theoretical_year_1990_2020} presents the
density function of AFI across occupations (not employment) in 1990 and
2020. It shows a clear increase in the age-friendliness of occupations with
a shift in mass from the lower to the upper tail. In total, nearly three quarters of occupations saw an increase in their AFI.

\begin{figure}[!t]
\begin{adjustbox}{max totalsize = {\textwidth}{0.3\textheight}, center}
		\includegraphics[width = \textwidth]{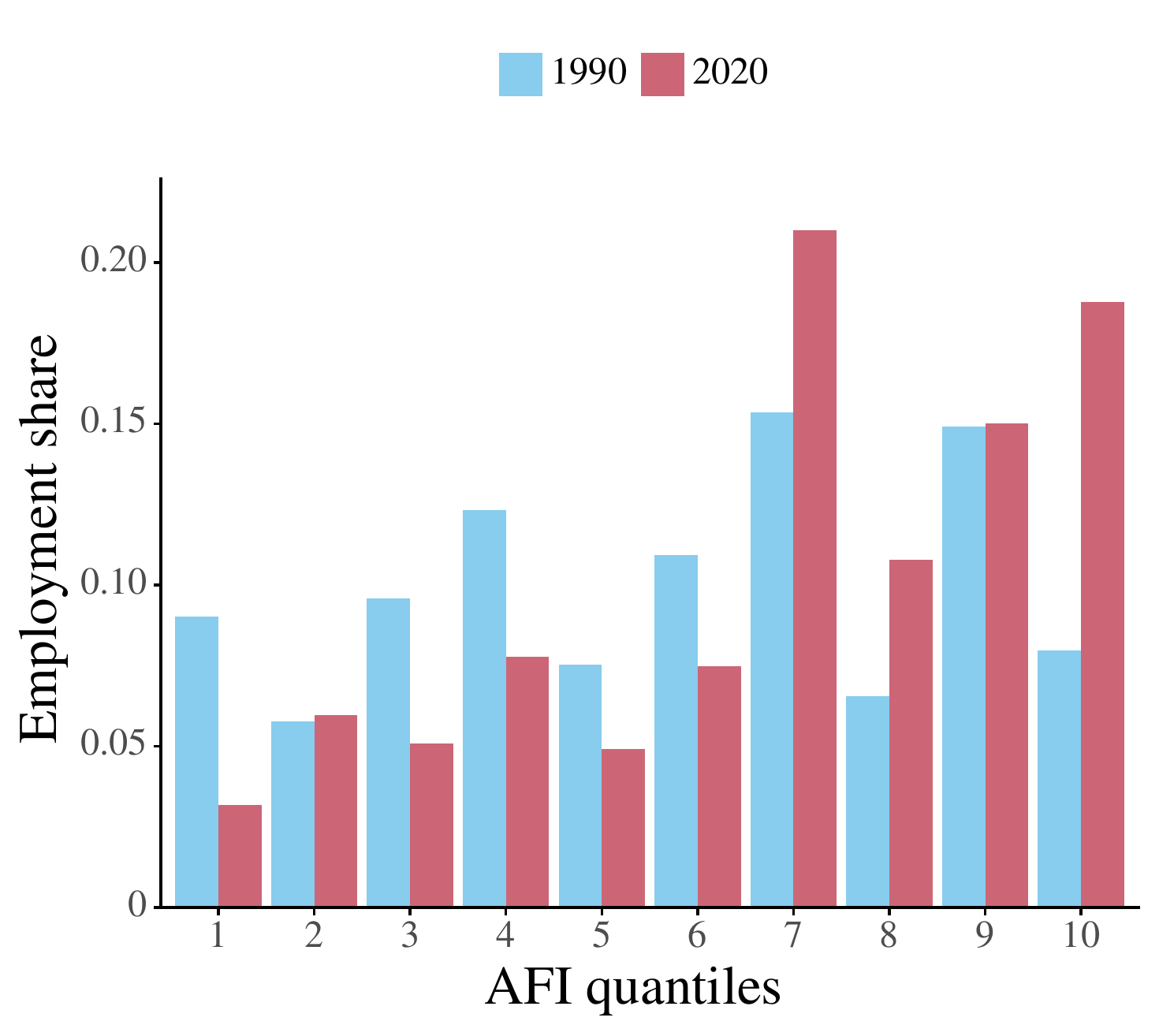}
	\end{adjustbox}
 \caption{Shifts in Employment Share by AFI Decile}
 \label{Fig:barchart_dodge_employment_shares_vs_maestas}\VerticalSpaceFloat
 {\footnotesize
    \textit{Notes:} This figure shows employment shares broken down by AFI deciles in 1990 and 2020. Deciles are defined by numerical values based on the 2020 AFI.
    \par}
\end{figure}

This rightward shift is reflected in Figure \ref{Fig:barchart_dodge_employment_shares_vs_maestas}, which depicts employment
shares by AFI decile (defined using the AFI in 2020). As a result of the
increase in age-friendliness in so many occupations, there are fewer
occupations in the lower deciles and more in the upper deciles of the AFI.
Consequently, the pattern revealed by Figure \ref{Fig:barchart_dodge_employment_shares_vs_maestas} is one of declining
employment shares in below average age-friendly occupations and an increase
for above average ones. Relatedly, between 1990 and 2020, there was an
increase of 49 million in employment in above-average age-friendly
occupations. This exceeds the total increase in employment over
this period, 41.5 million, and reveals significant opportunities for
U.S. workers to move into age-friendly occupations.

\begin{figure}[!t]
\begin{subfigure}{.475\textwidth}
		\begin{adjustbox}{max totalsize = {\textwidth}{0.9\textheight}, center}
			\includegraphics[width = \textwidth]{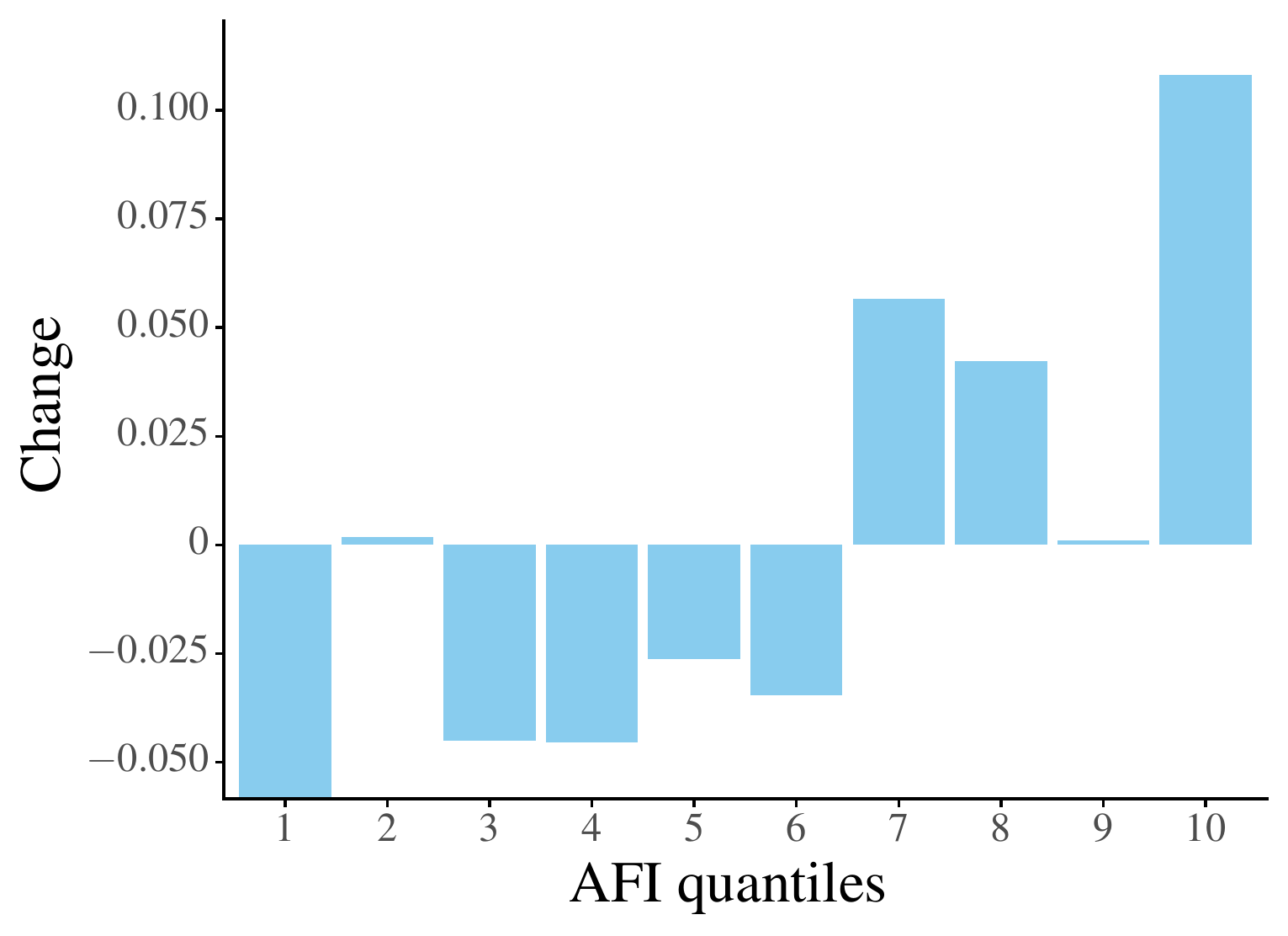}
		\end{adjustbox}	
		\caption{Change in employment shares}
		\label{Fig:barchart_dodge_change_share_by_afi_quantiles}
	\end{subfigure}\hfill 
\begin{subfigure}{.475\textwidth}
		\begin{adjustbox}{max totalsize = {\textwidth}{0.9\textheight}, center}
			\includegraphics[width = \textwidth]{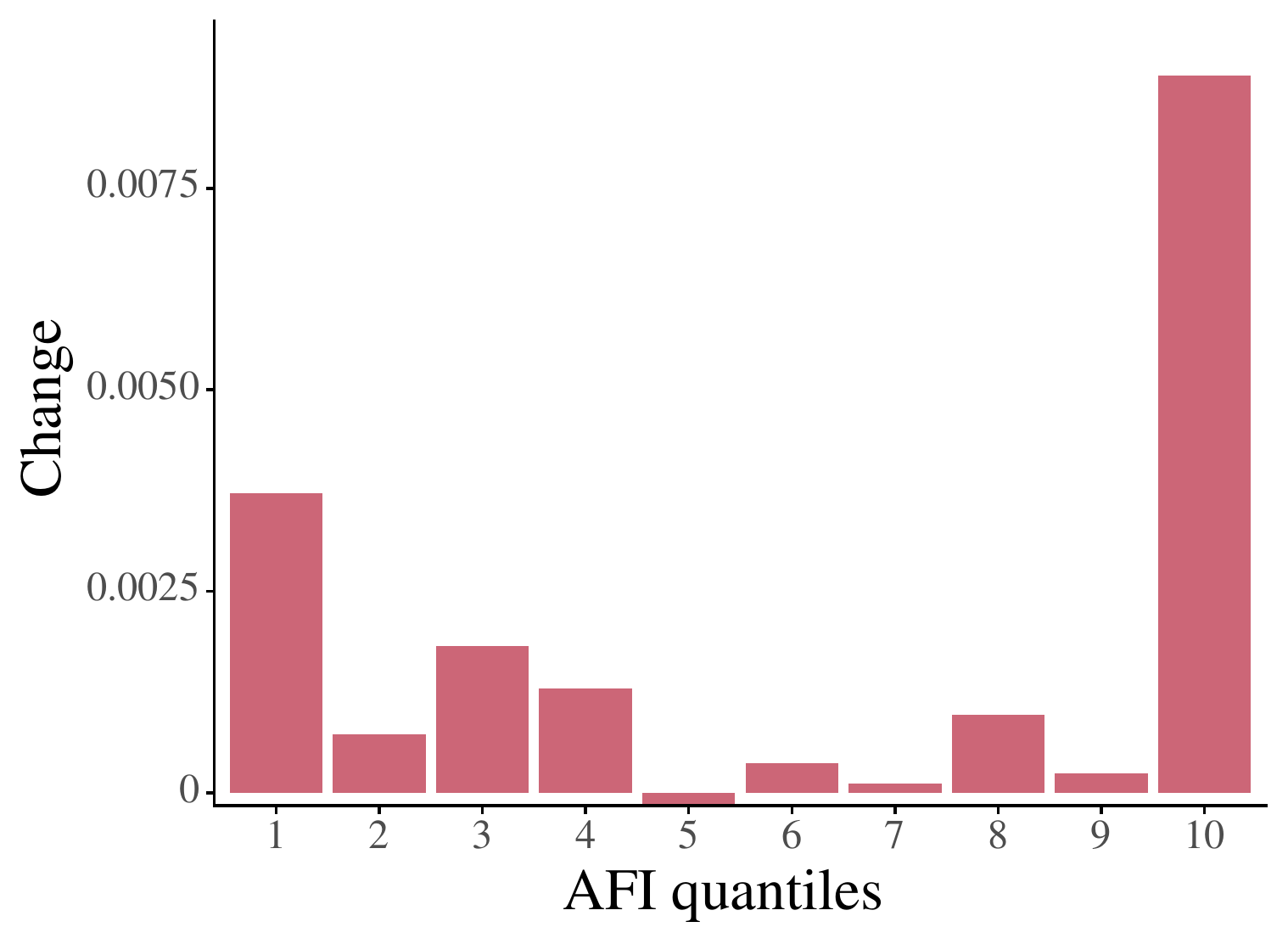}
		\end{adjustbox}	
		\caption{Change in AFI}
		\label{Fig:barchart_dodge_change_maestas_by_afi_quantiles}
	\end{subfigure}
\caption{Decomposition of in employment shares and AFI by AFI deciles}
\label{Fig:barchart_dodge_change_by_afi_quantiles}\VerticalSpaceFloat
{\footnotesize
    \textit{Notes:} This figures shows the percentage change in employment and AFI for each decile of the AFI index
    \par}
\end{figure}

Pursuant to these shifts, the average employment-weighted AFI across all
occupations increased by 8 per cent over this time period. It is important to
understand whether these substantive changes are driven by most occupations becoming more age friendly (an within occupations effect)
or by a shift towards more age-friendly types of occupations (an between occupations effect). We answer this
question by performing a \cite{Blinder1973}-\cite{Oaxaca1973} decomposition,
which reveals that the within component is much more important. In particular,
92 per cent of the rise in AFI is driven by within-occupation changes, depicted by the rightward shift in Figure \ref{Fig:density_maestas_theoretical_year_1990_2020}. There is also a between-occupations effect, which further boosts AFI as above-average age-friendly occupations saw employment increase by 44 per cent, compared to a 21
per cent increase for below-average occupations. Figure \ref{Fig:barchart_dodge_change_by_afi_quantiles} provides a further look at this
between-occupations effect by depicting the 1990 and 2020 employment shares by AFI quartiles and shows a sizable shift away from the lowest AFI quartiles. 

\begin{figure}[!t]
\begin{subfigure}{.475\textwidth}
		\begin{adjustbox}{max totalsize = {\textwidth}{0.9\textheight}, center}
			\includegraphics[width = \textwidth]{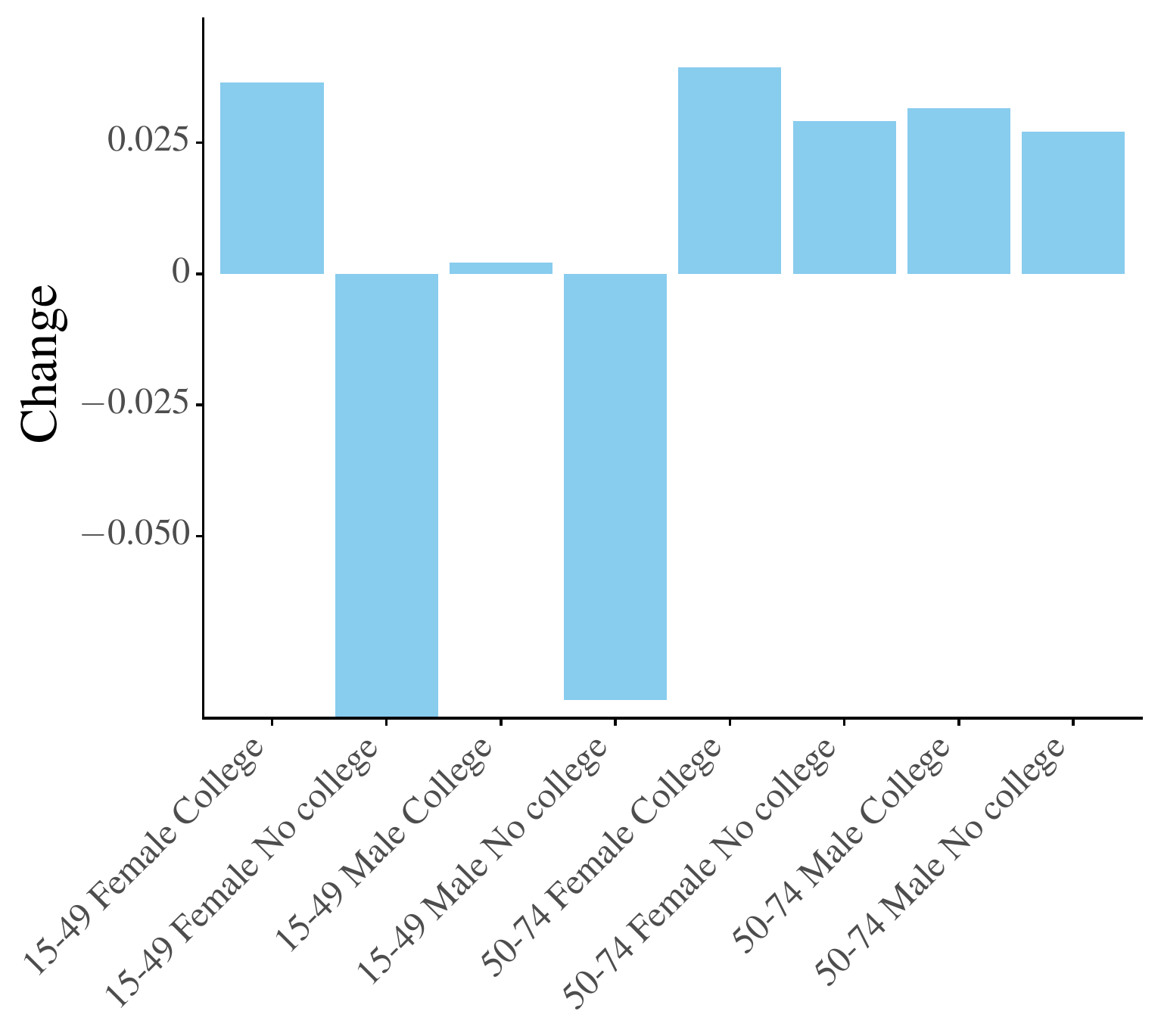}
		\end{adjustbox}	
		\caption{Change in employment shares}
		\label{Fig:barchart_dodge_change_share_by_demographic_groups}
	\end{subfigure}\hfill 
\begin{subfigure}{.475\textwidth}
		\begin{adjustbox}{max totalsize = {\textwidth}{0.9\textheight}, center}
			\includegraphics[width = \textwidth]{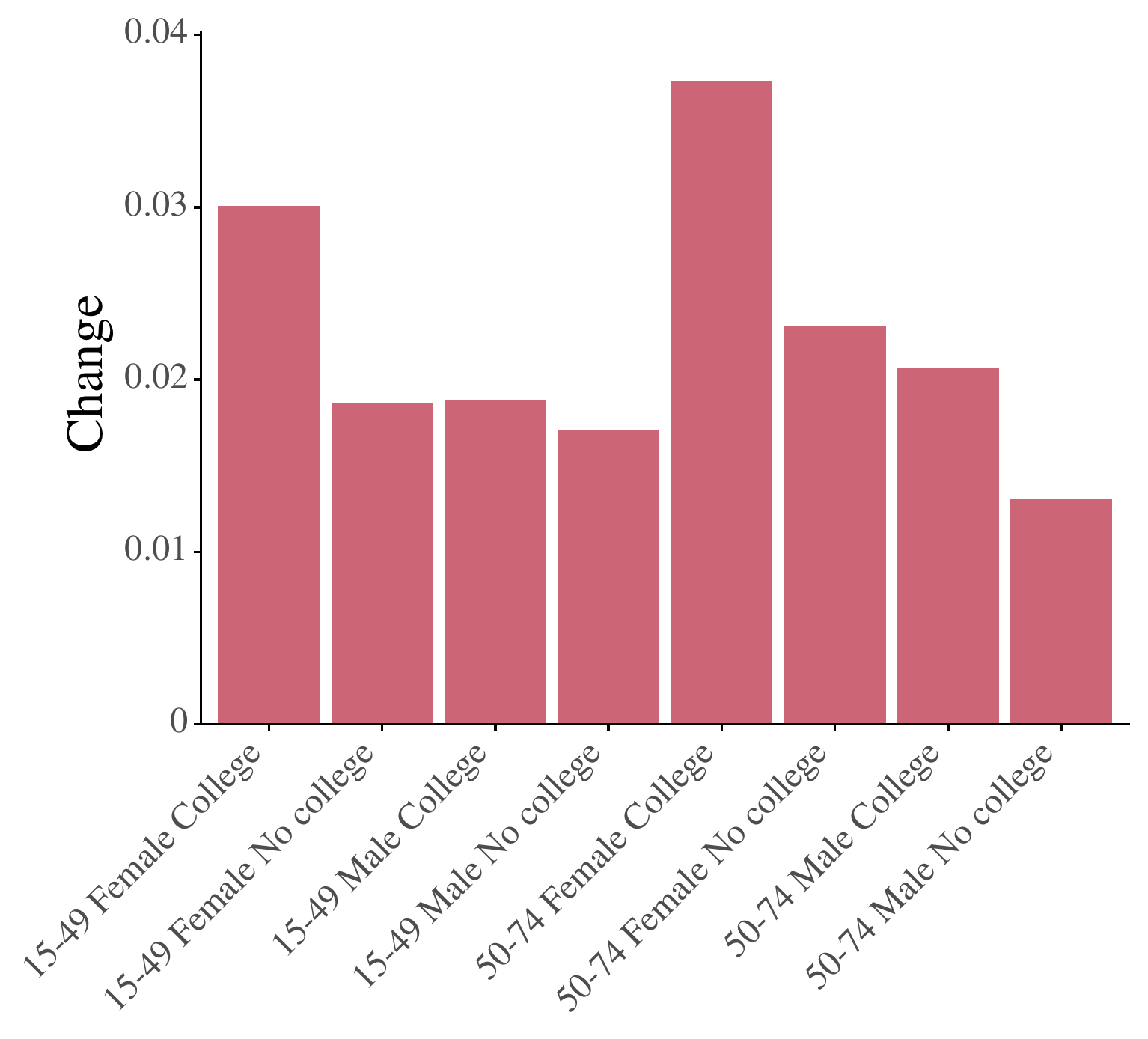}
		\end{adjustbox}	
		\caption{Change in AFI}
		\label{Fig:barchart_dodge_change_maestas_by_demographic_groups}
	\end{subfigure}
 \caption{\newline Changes in Employment Shares and AFI by demographic groups}
\label{Fig:barchart_dodge_change_by_demographic_groups}\VerticalSpaceFloat
 {\footnotesize 
    \textit{Notes:} This figures shows the percentage change in employment and employment weighted change in AFI for demographic groups governed by age, sex, and education.
    \par}
\end{figure}

Figure \ref{Fig:barchart_dodge_change_by_demographic_groups} turns to
changes in the U.S. labour market from the worker perspective. It shows a general shift towards older, female, and more educated workers. Figure \ref{Fig:barchart_dodge_change_share_by_demographic_groups} shows a significant decline in the employment shares of younger male and female non-college graduates, driven by absolute declines in their employment of 1.1 and 1.2 million, respectively. Mirroring this is a pronounced rise in graduate employment at younger ages, which is particularly noticeable for females. Younger female graduate employment rose by 9.1 million between 1990 and 2020, compared to only 4.8 million for men. More strikingly, Figure \ref{Fig:barchart_dodge_change_maestas_by_demographic_groups} reveals an increase in AFI for occupations held by all demographic groups, though the largest gains are amongst older workers, graduates, and females with the largest increase of all for older female graduates. This widespread increase in the age-friendliness of occupations across demographic groups signals a notable pattern, which we discuss in the rest of the paper. 

This trend is reflected in Table \ref{Tbl:ProbabilityTopQuartile}, which presents the employment shares of occupations in the top quartile of age-friendly occupations broken down by demographic group. Of the 33.1 million increase in employment in this top quartile of age-friendly occupations, only 15.2 million is accounted for by workers aged over 50 years, whereas female and college graduates account for most of the more age-friendly jobs that have been created since 1990. Note too that the increase for non-college, older males is quite small---from 20.3 per cent to only 25.9
per cent. Figure \ref{Fig:barchart_dodge_change_by_demographic_groups} and Table \ref{Tbl:ProbabilityTopQuartile} reveal that females and college graduates have been major beneficiaries of age-friendly jobs and the older workers who have consequently lost out the most have been male non-graduates. 

\begin{table}[!t]
\caption{Probability in a top quartile age-friendly job}
\begin{adjustbox}{max totalsize = {\textwidth}{0.6\textheight}, center}
		\begin{tabular}{lllrrr}
\toprule
           &      &       &   1990 &   2020  & Change \\
Education & Sex & Age &        &        &                                                                                    \\
\midrule
Total & Female & 15-49 & 31.4\% & 48.2\% & 16.8\\
           &      & 50-74 & 31.8\% & 51.5\% & 19.7 \\
           & Male & 15-49 & 20.6\% & 31.7\% & 11.1 \\
           &      & 50-74 & 23.0\% & 34.6\% & 11.6 \\
College & Total & 15-49 & 27.5\% & 48.5\%  & 21.0 \\
           &      & 50-74 & 28.1\% & 53.2\% & 25.1 \\
           & Female & 15-49 & 25.7\% & 52.2\% & 26.5 \\
           &      & 50-74 & 23.0\% & 56.2\% & 33.2 \\
           & Male & 15-49 & 29.0\% & 44.3\% & 15.3 \\
           &      & 50-74 & 30.7\% & 50.5\%  & 19.8 \\
No college & Total & 15-49 & 24.9\% & 34.8\%  & 9.9 \\
           &      & 50-74 & 26.5\% & 36.9\%  & 10.4 \\
           & Female & 15-49 & 33.2\% & 45.7\%  & 12.5 \\
           &      & 50-74 & 33.6\% & 49.0\%  & 15.4 \\
           & Male & 15-49 & 17.8\% & 26.0\%  & 8.2 \\
           &      & 50-74 & 20.3\% & 25.9\% & 5.6 \\
\bottomrule
\end{tabular}
	\end{adjustbox}\VerticalSpaceFloat
 \label{Tbl:ProbabilityTopQuartile}
 {\footnotesize 
     \textit{Notes:} This table shows proportion of different demographic groups employed in top quartile age-friendly occupations. Quartiles are defined by numerical values based on 2020 AFI.
    \par}
\end{table}

\subsection{Limits to Comparative Advantage?}

The patterns depicted so far, especially those in Figure \ref{Fig:barchart_dodge_change_maestas_by_demographic_groups}, are at first
puzzling. U.S. jobs have become significantly more age friendly and on
the basis of comparative advantage, we would have expected older workers to be the primary beneficiaries. Instead, it is younger college graduates and females who have been the ones moving into these age-friendly jobs.

\VerticalSpace
In principle, there are three reasons that could account for this pattern. Each is informative about the broader dynamics of the U.S. labour market and the direction of future research in this area.

\VerticalSpace
\textit{First}, our AFI may capture not just the age-friendliness of occupations, but also their \textquote{female-} and \textquote{graduate-friendliness}. A confirmation that this is the case, at least to some degree, comes from two pieces of evidence. To start with, Table \ref{Tbl:Regression_results} shows that while AFI significantly explains the share of older workers in an occupation, its explanatory power is limited, suggesting that other demographic groups are also typically employed in high AFI occupations. In addition, there is evidence that attributes that make jobs more age friendly---less physical exertion, greater use of social and communication skills, and less harsh environmental conditions---are also more attractive to female and more educated workers. This is confirmed in Table \ref{Tbl:onet_characteristics_by_all_mix_year2020_ajs}, which shows O*NET job
characteristics for younger female and younger college-educated workers as well as older workers. \cite{Maestasetal2018} also show that both females
and graduates are indeed more willing to pay for the occupational attributes valued by older workers than males and non-graduates, respectively. There may also be major differences between older college graduates and older non-college
workers in terms of their preferences and skills---a topic that has not
received much attention in the literature.

\begin{table}[!t]
\caption{Selected O*NET occupational characteristics by sex, education and age}
\begin{adjustbox}{max totalsize = {\textwidth}{0.6\textheight}, center}
		\begin{tabular}{lrrrrrrrrrr}
\toprule
Age & {50-74} & & & & & {15-49} & & & & \\
Sex &  All & Female &  & Male  & & Female & Male & All & All  \\
Education & All & College & No College & College & No College & All & All & College & No College \\
\midrule
Cognitive Abilities      & 0.45  &  0.485 &      0.427 &   0.491 &      0.430  & 0.447 & 0.444	& 0.487	& 0.423\\
Physical Abilities       & 0.168 &   0.112 &      0.161 &   0.106 &      0.236 & 0.158	& 0.204	& 0.113	& 0.219\\
Psychomotor Abilities    & 0.227 &   0.151 &      0.208 &   0.165 &      0.315 & 0.195 & 0.271	& 0.16	& 0.274\\
Sensory Abilities        & 0.295 &   0.279 &      0.270 &   0.289 &      0.330 & 0.274 & 0.312 & 0.283 & 0.3 \\
Work Output              & 0.362 &   0.313 &      0.334 &   0.335 &      0.427 & 0.331 & 0.397 & 0.327	& 0.387\\
Communication            & 0.696 &   0.773 &      0.687 &   0.762 &      0.631 & 0.707	& 0.663	& 0.761	& 0.643\\
Conflictual Contact      & 0.403 &   0.440 &      0.410 &   0.404 &      0.379 & 0.43  & 0.389 & 0.42  & 0.403 \\
Responsibility for Others & 0.563 &   0.538 &      0.527 &   0.561 &      0.608 & 0.536 & 0.584 & 0.54	& 0.572\\
Environmental Conditions  & 0.28 &   0.195 &      0.230 &   0.221 &      0.400 & 0.221 & 0.342 & 0.207	& 0.326 \\
Job Hazards               & 0.168 &  0.120 &      0.136 &   0.129 &      0.243 & 0.135	& 0.206 & 0.123	& 0.199\\
Pace and Scheduling       & 0.463 &  0.411 &      0.445 &   0.431 &      0.523 & 0.427 & 0.489	& 0.418 & 0.482 \\
Recognition               & 0.423 &  0.534 &      0.353 &   0.572 &      0.349 & 0.409	& 0.399	& 0.542	& 0.331\\
Working Conditions        & 0.498 &  0.597 &      0.431 &   0.625 &      0.441 & 0.475	& 0.475 & 0.601	& 0.409\\
\bottomrule
\end{tabular}
	\end{adjustbox}\VerticalSpaceFloat
 \label{Tbl:onet_characteristics_by_all_mix_year2020_ajs}
 {\footnotesize 
    \textit{Notes:} This table shows employment-weighted O*NET characteristics by sex, education and age in 2020.
    \par}
\end{table}

\textit{Second}, labour market allocations may not follow comparative advantage
because of various distortions and frictions in the assignment of workers to
jobs. One important factor is limited worker mobility, which is likely to increase with age as workers become settled into (permanent) jobs. Indeed, older workers have longer tenure in their jobs \citep{Allen2019}, either because general labour market imperfections have meant long-lasting matches or because these workers have accumulated firm-specific human capital, making them reluctant to leave these jobs and employers unwilling to replace them with others. In addition, with a shorter horizon until retirement, the return on learning will be diminished and the impact of fixed costs associated with occupational transitions rise especially in the presence of age discrimination in hiring \citep{Neumark2019}. Older workers may thus be understandably reluctant to change occupations. As emphasised by \cite{Autor2019}, these effects are also likely to produce higher median ages in \textquote{old job} occupations. Due to these inertial forces, the most rapidly growing occupations are likely to have more younger workers.

\begin{figure}[!t]
\begin{subfigure}{.475\textwidth}
		\begin{adjustbox}{max totalsize = {\textwidth}{0.9\textheight}, center}
			\includegraphics[width = \textwidth]{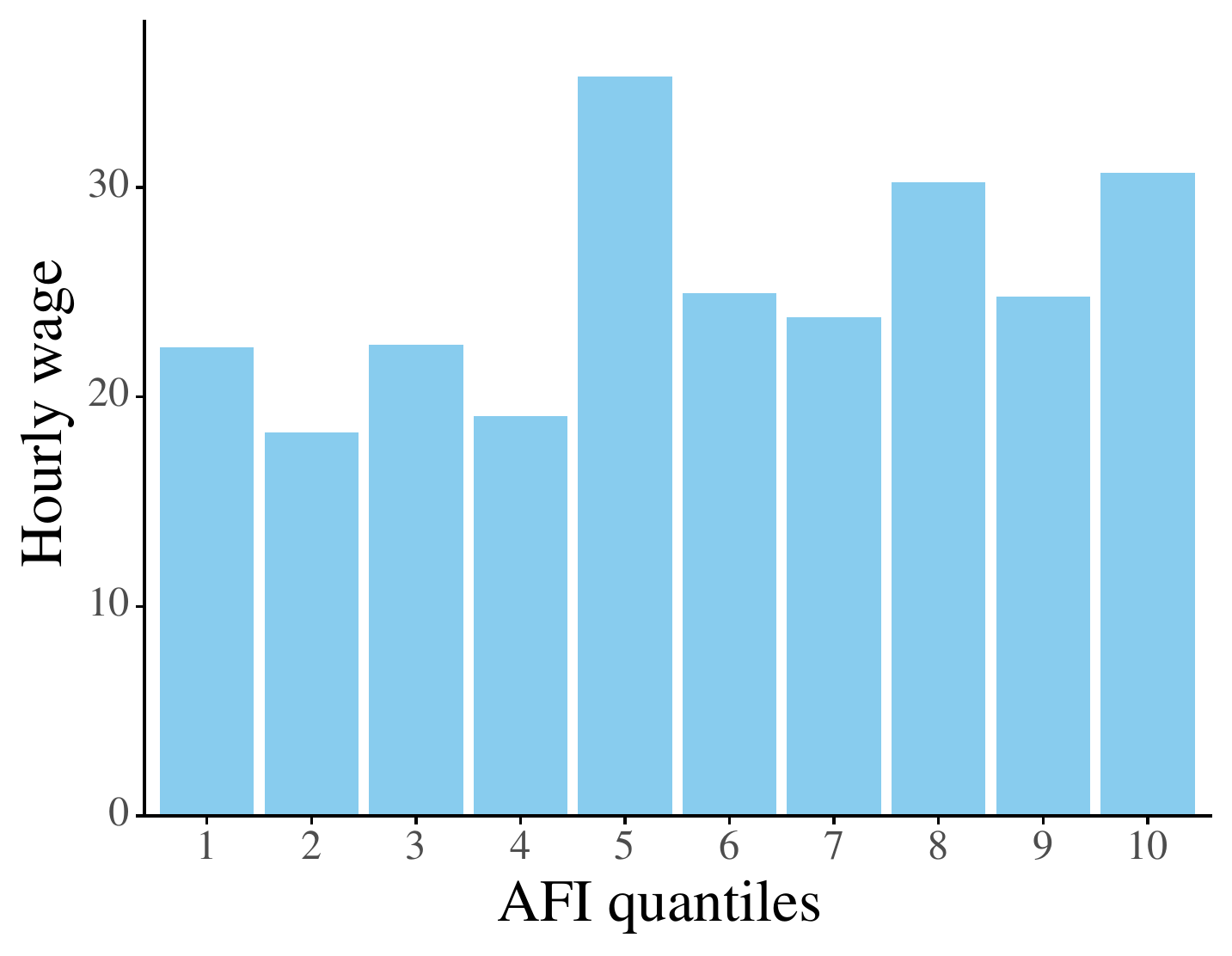}
		\end{adjustbox}	
		\caption{Hourly wage}
		\label{Fig:barchart_dodge_hourly_wage_by_afi_quantiles}
	\end{subfigure}\hfill 
\begin{subfigure}{.475\textwidth}
		\begin{adjustbox}{max totalsize = {\textwidth}{0.9\textheight}, center}
			\includegraphics[width = \textwidth]{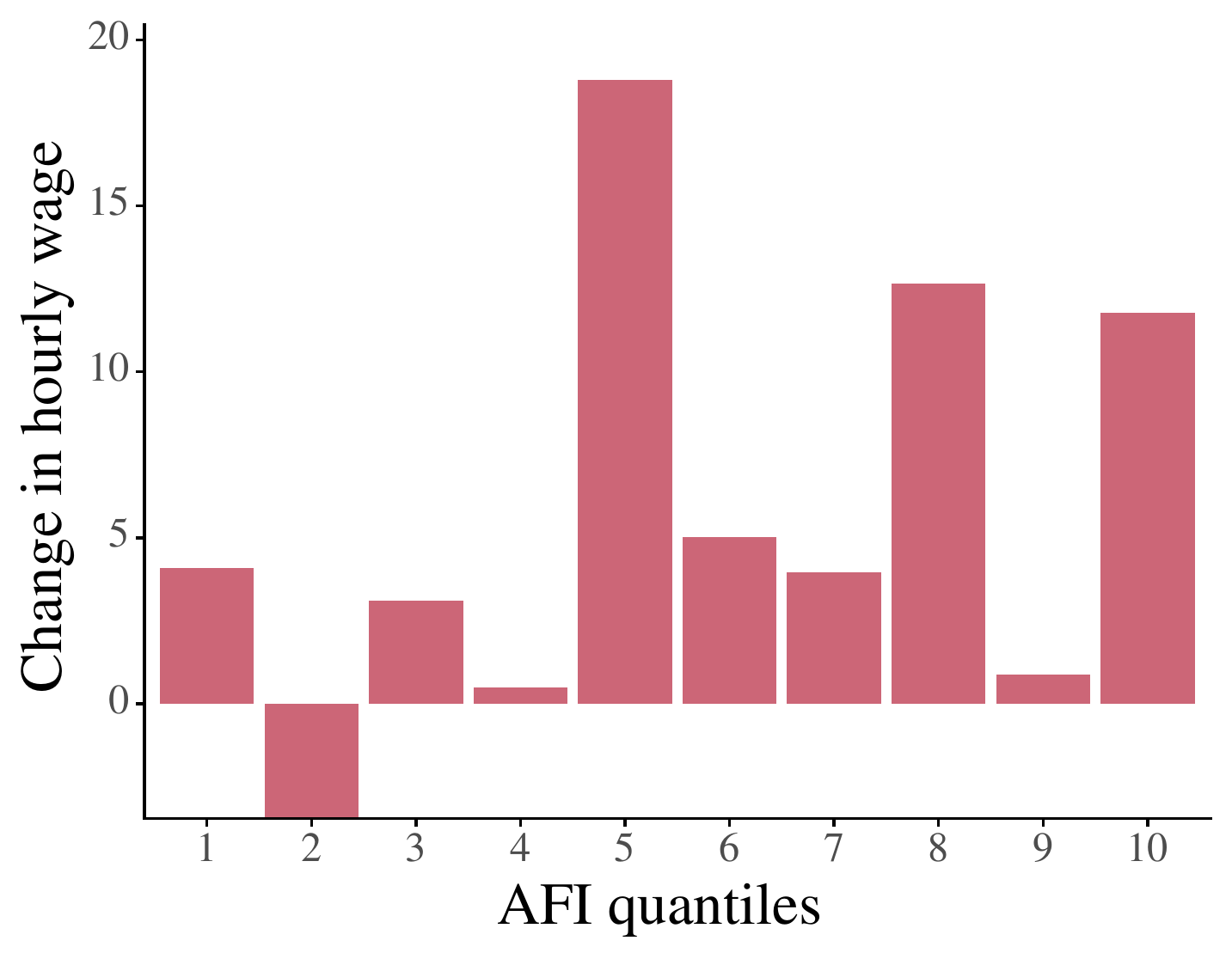}
		\end{adjustbox}	
		\caption{Change in hourly wage 1990-2020}
		\label{Fig:barchart_dodge_change_in_hourly_wage_by_afi_quantiles}
	\end{subfigure}
\begin{subfigure}{.475\textwidth}
		\begin{adjustbox}{max totalsize = {\textwidth}{0.9\textheight}, center}
			\includegraphics[width = \textwidth]{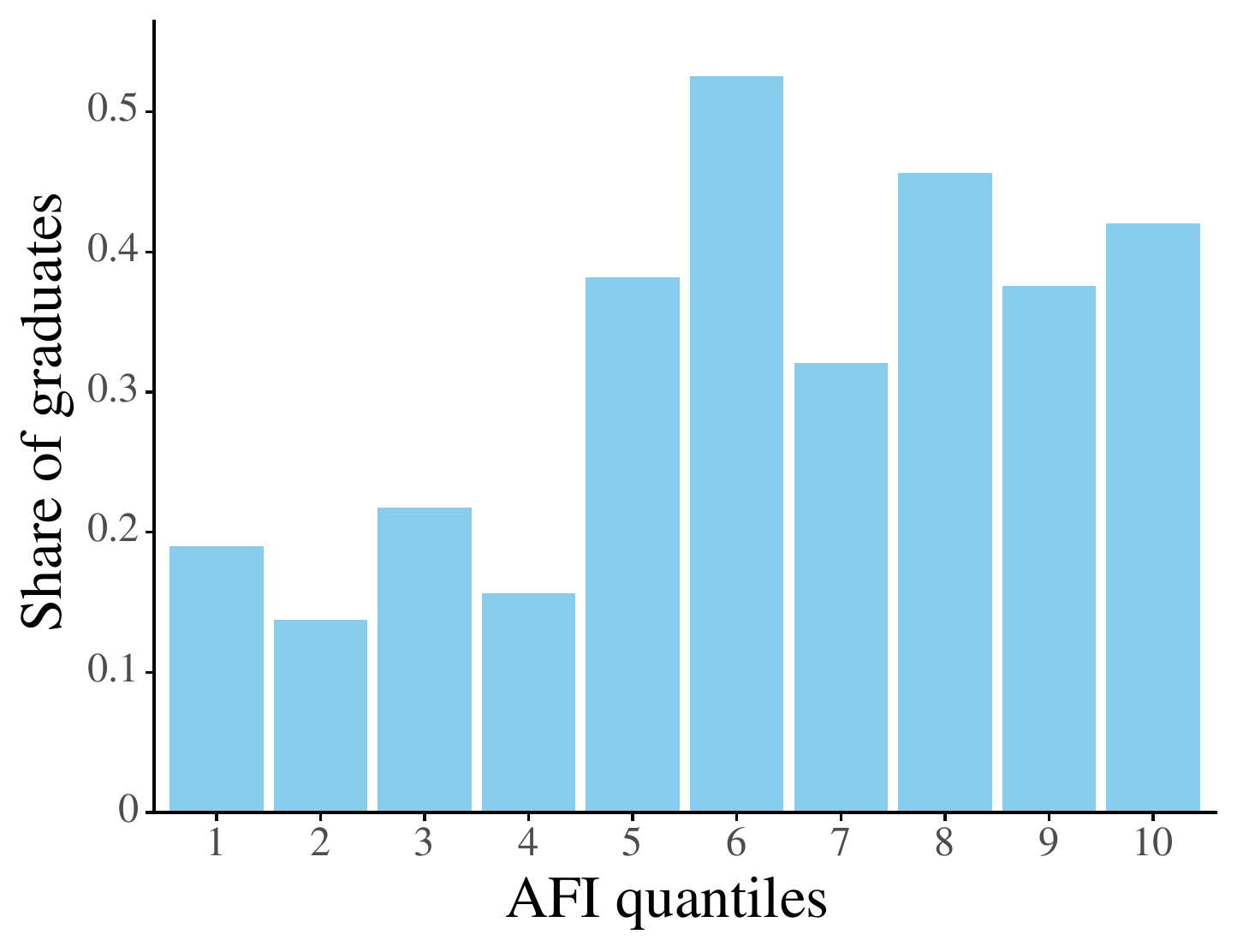}
		\end{adjustbox}	
		\caption{Share of graduates}
		\label{Fig:barchart_dodge_share_of_graduates_by_afi_quantiles}
	\end{subfigure}\hfill 
\begin{subfigure}{.475\textwidth}
		\begin{adjustbox}{max totalsize = {\textwidth}{0.9\textheight}, center}
			\includegraphics[width = \textwidth]{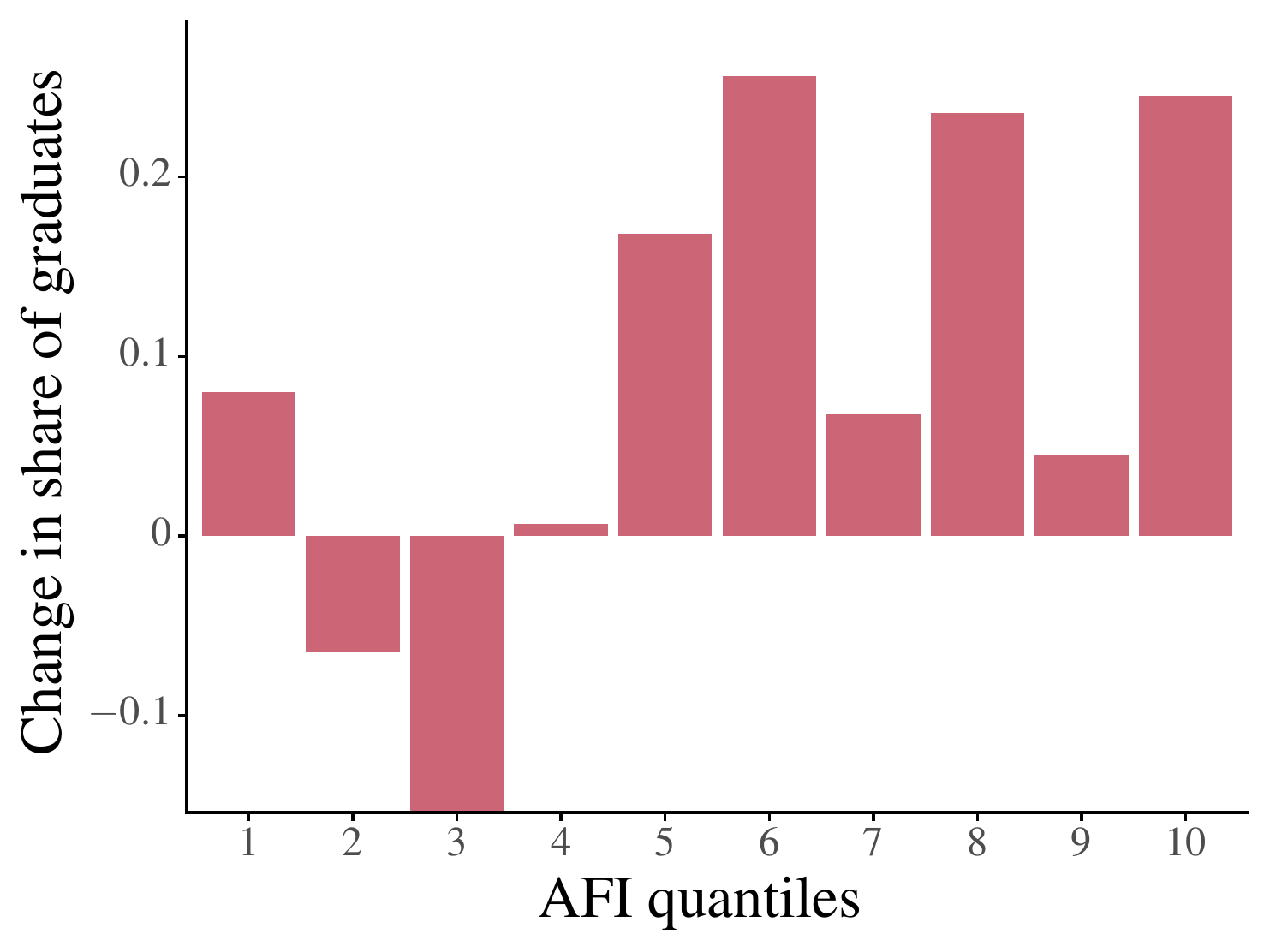}
		\end{adjustbox}	
		\caption{Change in share of graduates 1990-2020}
		\label{Fig:barchart_dodge_change_in_share_of_graduates_by_afi_quantiles}
	\end{subfigure}
\begin{subfigure}{.475\textwidth}
		\begin{adjustbox}{max totalsize = {\textwidth}{0.9\textheight}, center}
			\includegraphics[width = \textwidth]{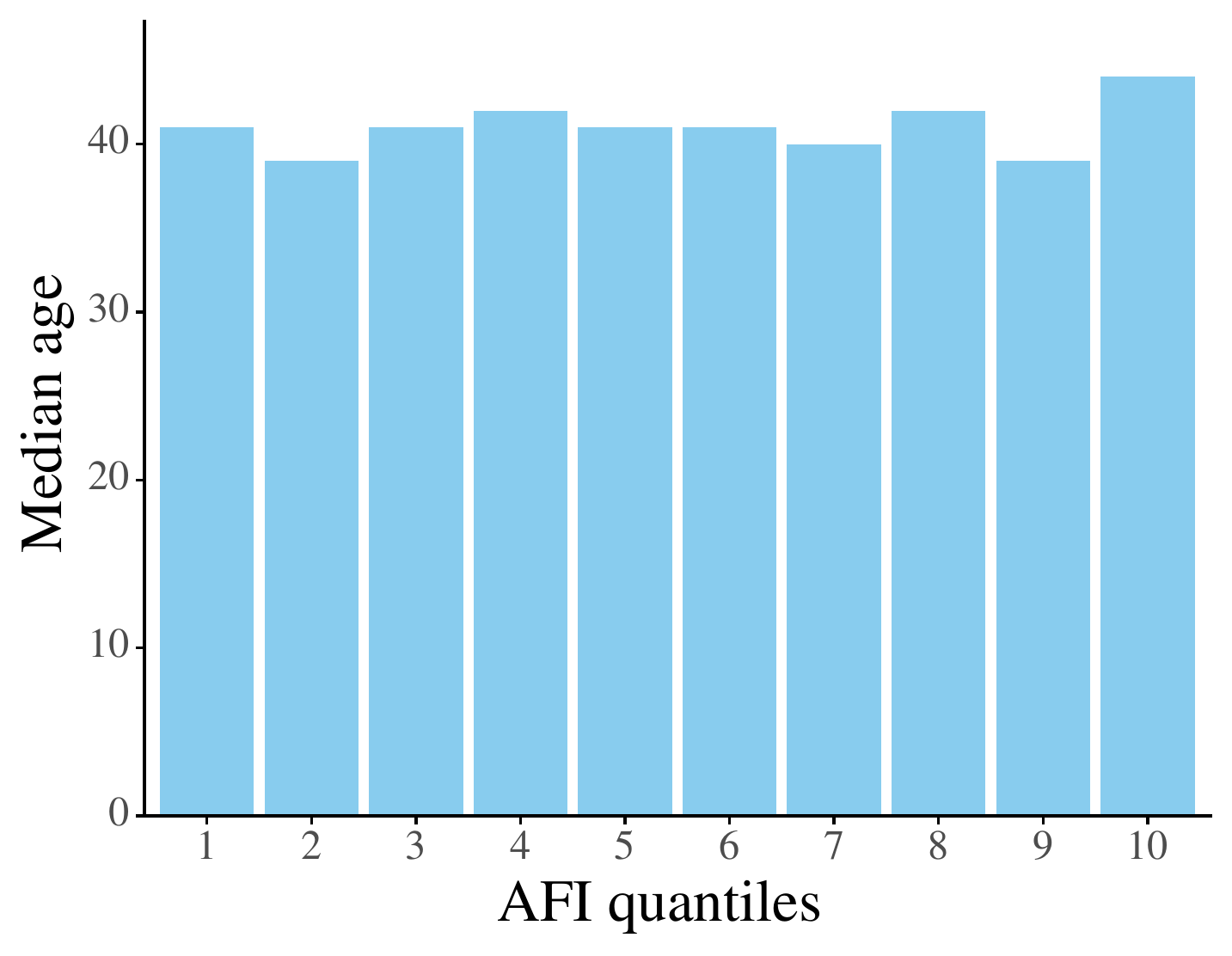}
		\end{adjustbox}	
		\caption{Median age}
		\label{Fig:barchart_dodge_median_age_by_afi_quantiles}
	\end{subfigure}\hfill 
\begin{subfigure}{.475\textwidth}
		\begin{adjustbox}{max totalsize = {\textwidth}{0.9\textheight}, center}
			\includegraphics[width = \textwidth]{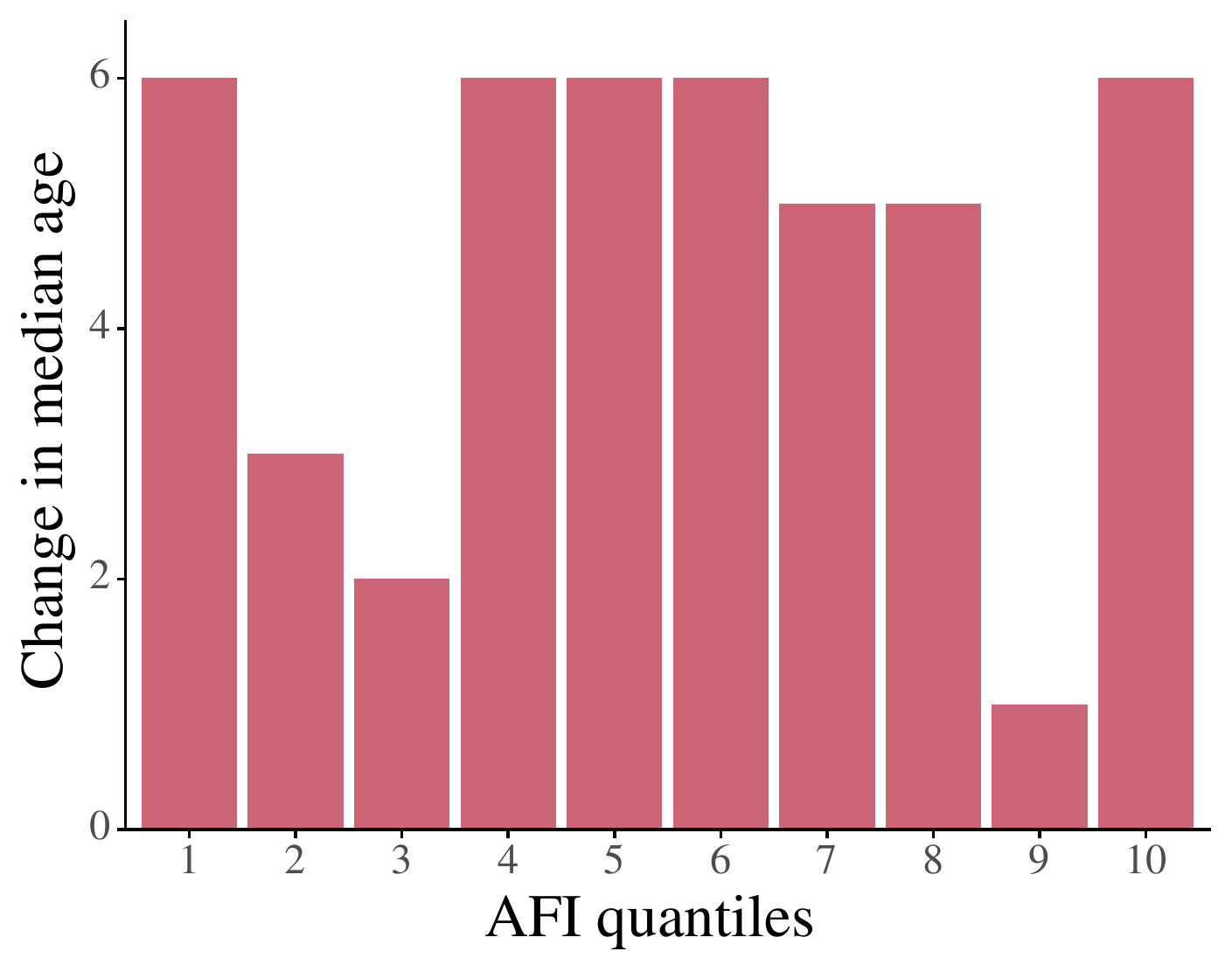}
		\end{adjustbox}	
		\caption{Change in median age 1990-2020}
		\label{Fig:barchart_dodge_change_in_median_age_by_afi_quantiles}
	\end{subfigure}
\caption{Comparing median age, wage, and education by AFI quantiles} 
\label{Fig:comparing_age_wage_educ_by_afi}\VerticalSpaceFloat
{\footnotesize 
    \textit{Notes:} This figure shows the hourly wage, share of graduates and median age in levels (LHS) and changes from 1990 to 2020 (RHS) by AFI deciles.
    \par}
\end{figure}

\VerticalSpace
\textit{Third}, when there is rent sharing or other frictions, employers may prefer higher-productivity workers, regardless of comparative advantage \citep{Acemoglu1999,ShimerSmith2000}. If younger college graduates and especially female college graduates have higher productivity (or are perceived to have higher productivity), then employers with new age-friendly jobs may give them priority ahead of older workers. Figure \ref{Fig:comparing_age_wage_educ_by_afi} provides suggestive evidence for this
explanation. The top and middle panels show that, in general, age-friendly
jobs have higher wages and have experienced higher wage growth \footnote{There is also evidence of polarisation whereby the lowest wage jobs have also not seen much benefit in terms of age friendliness}. It also
confirms that they have attracted a greater fraction of graduates. It also
provides another aspect of the puzzling patterns we have described so far:
median age has not changed much differentially across AFI quartiles.

\VerticalSpace
These potential market imperfections also shed light on which category of older workers have been most adversely affected in terms of missing out on age-friendly jobs. Older male non-graduates are the only category of older workers whose share of growth in top quartile jobs is below their share of growth in total employment (16 per cent and 27 per cent, respectively). Evidence for this can be seen in Table \ref{Tbl:onet_characteristics_by_all_mix_year2020_ajs}, which shows that compared to other older workers, non-graduate males are more likely to be in physical jobs (physical, psychomotor, and sensory abilities and work output), have more difficult work conditions (environmental conditions and job hazards) with less autonomy (responsibility for others) and less flexibility (pace and scheduling). 

\VerticalSpace
The reason for this difference is linked to the industries they work in. The largest number are in manufacturing (18.2 per cent) and construction (14.7 per cent), which rank at the bottom in terms of age-friendliness in Figure \ref{Fig:average_maestas_by_industry_year2020}. Whilst both industries have seen an improvement in their age-friendliness, the gap between them and the most age-friendly sectors has widened. By contrast, older female non-graduates are disproportionately in more age-friendly industries (38 per cent in professional and related services and 17 per cent in retail). 

\VerticalSpace
From the perspective of our three highlighted labour market imperfections, these patterns are revealing. They are consistent with the findings of \cite{Autor2009} that these industries disproportionately constitute \textquote{old jobs}. Another potential explanation is that we have so far focused on differences in preferences between young and old but it may be that younger and older graduates have more similar preferences than, say, older non-graduates and older graduates. Finally, given the fact that older graduates have benefited from age-friendly jobs, it is possible that different categories of older workers have different abilities to share the economic rents arising from absolute as well as comparative advantage. It is noticeable that older graduates in particular have benefited more from age-friendly jobs and so have gained from higher employment in higher-wage and more age-friendly occupations. 

\VerticalSpace
The disproportionately low access to age-friendly occupations for male non-graduates is a significant problem. This group have seen a substantial rise in labour force participation rates, which combined with relative cohort sizes means they represent the largest increase in older employment (7.6 out of 27.5 millions) but they have seen the fewest benefits from age-friendly jobs. This is also the group with the worst health conditions and the group that has experienced the worst declines in health \citep{zajacovaa2017} and so is most in need of age-friendly jobs. From the perspective of age-friendly jobs, these results suggest targeting policies on male non-graduates and on manufacturing and construction in particular combined with policies aimed at supporting retraining and aiding transitions into newer more age-friendly occupations \citep{AitkenSingh2022}. 

\VerticalSpace
Although a full exploration of the relative contribution of different labour market imperfections in influencing the distribution of age-friendly jobs is beyond this paper, our empirical results point to an important policy caveat. Policies aimed at improving labour market conditions for
older workers by encouraging greater age-friendliness may be much less
effective than previously presumed as these jobs are attractive to a wide range of demographic groups, especially females and graduates. Labour market imperfections interact with heterogeneity between older workers and homogeneity between specific categories of older and younger workers. The consequence is that purely age-based policies may have limited impact. A more holistic approach to policy taking into account not just the types of jobs for which older workers have skills and preferences, but also competition from other groups for the same jobs and the difficulties of transitioning these workers to new jobs may be necessary.

\section{Conclusion}\label{sec:conclusion}

The U.S. has witnessed a sharp, arguably epochal increase in employment
among older workers. Americans aged between 50-74 years used to make up 20 per cent of total employment in 1990 and now account for 33 per cent. How has this been achieved? What lessons does this hold for the future?

\VerticalSpace
To provide a first look as to how these substantive changes have been
achieved, we developed an age-friendliness index (AFI) using natural language processing methods and information on preferences and skills of older workers and detailed job characteristics. This index captures various aspects that are associated with preferences and skills of older workers. It identifies jobs that are performed in offices and less harsh environments, that do not involve high levels of physical exertion, and rely on social and communication skills as
more age friendly.

\VerticalSpace
After documenting that AFI aligns well with responses from a small-scale
survey and predicts the older workers' share of occupations in
1990, we study whether U.S. jobs have become more age friendly. We uncover
two major facts.

\VerticalSpace
First, U.S. jobs have become significantly more age friendly. This is driven
mostly by all occupations moving in an age-friendlier direction with only a small part played by a shift in employment away from low AFI towards higher AFI
occupations.

\VerticalSpace
Second, most of the age-friendlier jobs have not been taken up by older
workers, but by females and college graduates. Amongst older workers, females and graduates have also benefited but the biggest excluded group have been non-college males. We have discussed several reasons why this is---most importantly,
the overlap between characteristics that make jobs similarly attractive to females and college graduates as well as older workers; the unwillingness or inability of older workers to move away from their existing jobs; and the possible preference of employers in an imperfect labour markets to hire younger, more productive college-graduate workers rather than older non-college workers for these jobs.

\VerticalSpace
Two conclusions emerge from these empirical findings. The first is that due to the overlap in occupational characteristics between older and younger workers, the creation of age-friendly jobs is effectively part of a broader policy of creating good jobs. Given that an ageing society requires an increase in labour force participation at all ages, this is not necessarily a problem but it does reduce the ability of age-friendly jobs to boost employment at older ages whilst minimising the labour market impact on other groups. Second, the interaction of age-friendly policies with labour market imperfections means that different older workers will be impacted very differently. The manner in which older graduates have benefited from age-friendly jobs compared to non-graduates, especially males, suggests the value of supplementing general age-based policies with more targeted approaches. For older male non-graduates that would likely be either a focused improvement in the age friendliness of specific occupations or aiding transitions to more age-friendly occupations. 

\clearpage
\phantomsection
\addcontentsline{toc}{section}{References}
\bibliographystyle{ecta}
\bibliography{refs.bib}

\newpage 
\appendix

\FloatBarrier
\clearpage
\section*{Appendix}\label{sec:Appendix}
\subsection{Data}\label{App:Data}
\paragraph*{Census}
To generate the data, we follow \cite{Acemoglu2011,Autor2019,Acemoglu2022} among others in using the US Census of Population data for 1990, and the pooled American Community Survey (ACS) data for 2020, sourced from IPUMS \citep{Ruggles2018}. This gives us employment numbers by year, age, industry, sex, education, and occupation.

\begin{figure}[!t]
\begin{adjustbox}{max totalsize = {\textwidth}{0.3\textheight}, center}
    \includegraphics[width = \textwidth]{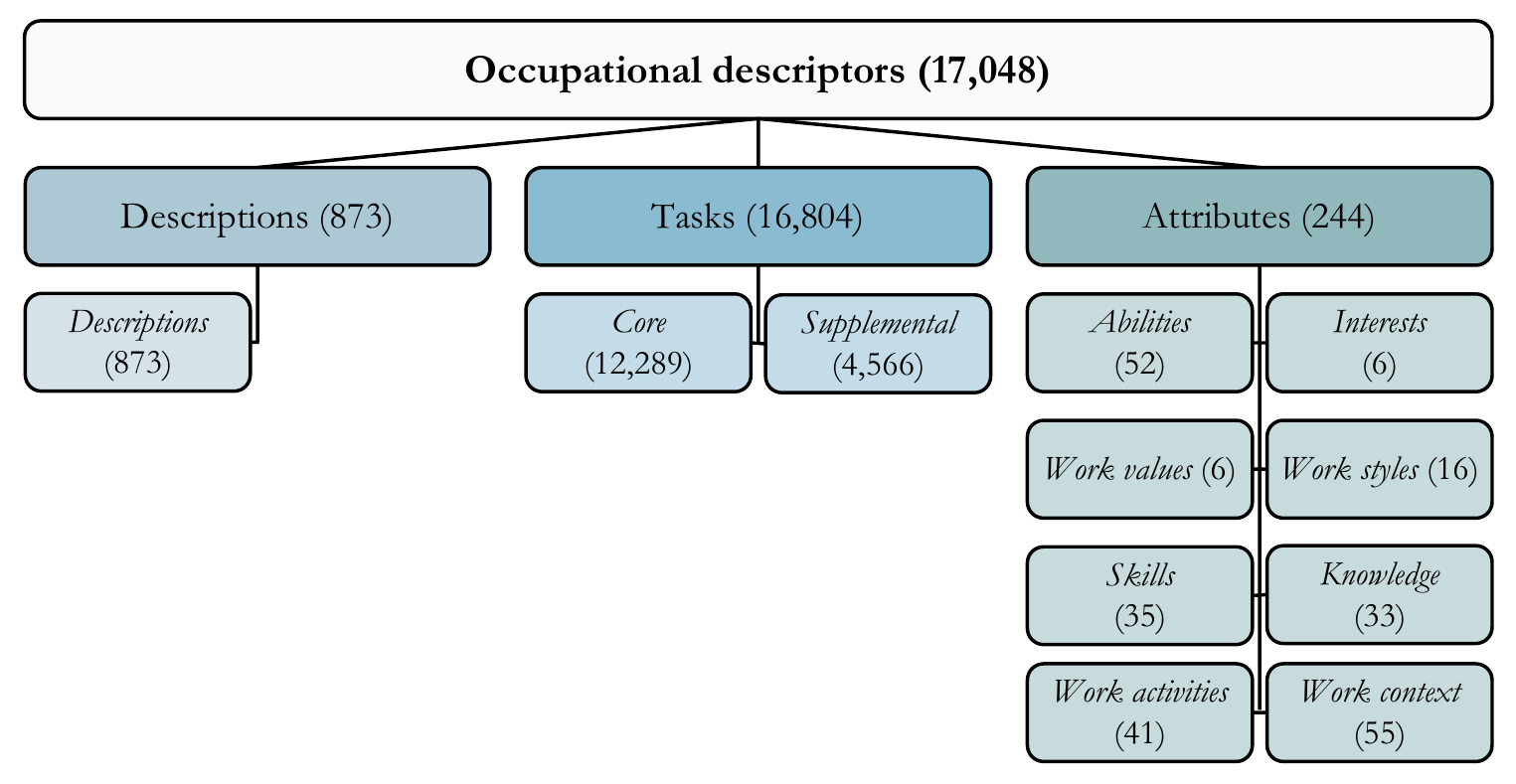}
\end{adjustbox}	
\caption{Occupational descriptors and subcategories}
\label{Fig:onet_data}\VerticalSpaceFloat
{\footnotesize
\textit{Notes:} This figure shows the occupational descriptors provided by O*NET that can be expressed meaningfully by both text and numeric scores. The 17,048 unique occupational descriptors consist of 873 occupation descriptions, 16,804 detailed tasks, and 244 attributes. Note that a small number of tasks may be core to some occupations while supplemental to others, meaning that the sum of core and supplemental tasks exceed the total number of unique tasks. We divide the descriptors into ten categories of which one category belongs to descriptions, one category belongs to tasks (core and supplemental tasks are grouped as one category), and eight categories belong to attributes. The figure is adapted from \cite{muhlbach2021} by courtesy of the author.
\par}
\end{figure}

\paragraph*{O*NET}
Occupational Information Network (O*NET) is a detailed database, providing the most comprehensive occupational information to our knowledge. O*NET keeps track of hundreds of descriptors for more than a thousand standardized occupations. The descriptors cover four broader types of information about an occupation, which may be decomposed into ten detailed categories, namely, \textit{job descriptions}, \textit{tasks}, \textit{abilities}, \textit{interests}, \textit{work values}, \textit{work styles}, \textit{skills}, \textit{knowledge}, \textit{work activities}, and \textit{work context}. Figure \ref{Fig:onet_data} displays an overview of the data available from O*NET.

O*NET updates the database multiple times a year. To obtain the data from 2020, we download the databases 24.2 (February 2020), 24.3 (May 2020), 25.0 (August 2020), and 25.1 (November 2020) and we use the average per descriptor by occupation. The data was assessed May 5, 2022 from this \href{https://www.onetcenter.org/db_releases.html}{link}. Regarding 1990, we use the 47 databases available between 2002 and 2022 and extrapolate using the Piece-wise Cubic Hermite Interpolating Polynomial (PCHIP) algorithm \citep{Fritsch1984}. The PCHIP algorithm is chosen as it outperforms several alternative algorithms in a pseudo out-of-sample validation exercise. We restrict the extrapolation such that all data points lie within one standard error of the linear extrapolation.

Cognitive Abilities - Abilities that influence the acquisition and application of knowledge in problem solving
Physical Abilities - Abilities that influence strength, endurance, flexibility, balance and coordination
Psychomotor Abilities - Abilities that influence the capacity to manipulate and control objects
Sensory Abilities - Abilities that influence visual, auditory and speech perception
Work Output - What physical activities are performed, what equipment and vehciles are operated/controlled, and what complex/technical activities are accomplished as job outputs?
Communication - Types and frequency of interactions with other people that are required as part of this job
Conflictual Contact - Amount of conflict that the workers will encounter as part of this job
Responsibility for Others - Amount of responsibility the worker has for other workers as part of this job
Environmental Conditions - Description of extreme environmental conditions the worker will be placed in as past of this job
Job Hazards -Descriptions of types of hazardous conditions the worker could be exposted to as pasrt of this job
Pace and Scheduling -Description of the role that time plays in the way the worker performs the tasks required by this job
Recognition - Occupations that satisfy thus work value offer advancement, potential for leadership, and are often considered prestigious
Working Conditions - Occupations that satisfy this work value offer job security and good working conditions. 

\subsection{Definitions of job amenities}\label{App:Definitions}
In the following, we provide several textual definitions of various job amenities (Definitions \ref{Def:amenity_schedule_flexibility}-\ref{Def:amenity_meaningful_work}) that are used to construct of index of age-friendliness in accordance with \cite{Maestasetal2018}. We also provide our definition of age-friendliness that is used for external validation in Definition \ref{Def:agefriendliness_survey}. The external validation is explained in Appendix \ref{App:Validation} below.

\begin{definition}[Schedule flexibility]\label{Def:amenity_schedule_flexibility}
A flexible work schedule allows employees a level of autonomy to create their own schedules and find a work-life balance that works for them. A flexible schedule allows employees to plan, vary, and adapt the times they begin and end their workday and to have some control of the working hours.
\end{definition}

\begin{definition}[Telecommuting]\label{Def:amenity_telecommuting}
Telecommuting is the ability of an employee to complete work assignments from outside the traditional workplace by using telecommunications tools such as email, phone, chat, and video apps. Often this means working from home or at a location close to home, such as a coffee shop, library, or co-working space.
\end{definition}

\begin{definition}[Physical job demands]\label{Def:amenity_physical_job_demands}
Physical demands refer to the level and duration of physical exertion generally required to perform job tasks, such as sitting, standing, carrying, walking, climbing stairs, lifting, carrying, reaching, pushing, and pulling, and it also includes strength, flexibility, dexterity, vision, and endurance.
\end{definition}

\begin{definition}[Work pace]\label{Def:amenity_work_pace}
Work pace is the rate at which an employee completes tasks and duties at the job.
\end{definition}

\begin{definition}[Work autonomy]\label{Def:amenity_work_autonomy}
Work autonomy is the degree to which the job provides substantial independence and discretion to the individual in scheduling the work and in determining the procedures to be used in carrying it out. Autonomy at work thus refers to how much freedom employees have to do their jobs.
\end{definition}

\begin{definition}[Paid Time Off]\label{Def:amenity_pto}
Paid time off (PTO) refers to the time that employees are paid for when they are not working. PTO includes paid vacation, sick time, holidays, and personal days.
\end{definition}

\begin{definition}[Teamwork]\label{Def:amenity_teamwork}
Working in teams means working with a group of people to achieve a shared goal or outcome effectively, listening to other members of the team, working for the good of the group as a whole, and having a say and sharing responsibility.
\end{definition}

\begin{definition}[Job training]\label{Def:amenity_job_training}
Job training means any type of instruction or a program for skill development and competence acquisition provided by the workplace. Job training provides opportunities to gain valuable new skills and enables career advancement.
\end{definition}

\begin{definition}[Meaningful work]\label{Def:amenity_meaningful_work}
Meaningful work refers to feeling morally, socially, personally, and spiritually significant and helps people feel a part of something larger than themselves, including being part of a community or society. Meaningful work contributes to the feeling of a purpose in life.
\end{definition}

\begin{definition}[Age-friendliness (for surveys)]\label{Def:agefriendliness_survey}
An age-friendly job appeals to older workers in particular. This will depend on a variety of characteristics including the following:
\begin{enumerate}[(a)]
  \item Should not involve intense or demanding physical work
  \item Should not involve high stress levels, such as tight deadlines, performance assessment, etc.
  \item Should encourage older workers to use their softer skills e.g., working in teams, dealing with interpersonal issues, etc.
  \item Should offer the opportunity for flexible working including part-time and variable hours
  \item Should offer autonomy and discretion rather than close management and supervision
  \item Provide an environment that is inclusive and supportive of older workers and not one where older workers are vulnerable to discrimination and abuse
\end{enumerate}
\end{definition}

\subsection{External validation through surveys}\label{App:Validation}
We distributed the surveys via three channels. The first channel is Amazon Mechanical Turk (MTurk), a crowd-sourcing marketplace for tasks that require human intelligence to complete. The use of MTurk in experimental research has been extensively validated \citep{Gabriele2010,berinsky_huber_lenz_2012,woo_keith_thornton_2015} and the service has been used by, e.g., \cite{Capraro2018}. We recruit workers from MTurk who are based in the US and who have graduated college. These participants were paid for their contribution. The second channel was students at London Business School participating in either an elective course "The Business Of Longevity" or core macroeconomic classes on Executive MBA or Sloan programmes. As a third channel to recruit participants, we distribute the surveys with anonymous links via social media, specifically Twitter and LinkedIn. We recruited 112 participants from MTurk, 58 from students, and 40 via social media, totaling 210 survey participants. We pool the responses from all survey participants and analyze them jointly, but our findings hold for each channel separately (results for each are available upon request).

\VerticalSpace
As described in the paper, participants were asked to score 10 occupations, each drawn from a different decile of our age-friendliness ranking. They were provided with Definition \ref{Def:agefriendliness_survey} of what makes an occupation \textquote{age friendly}.

\VerticalSpace
Scoring was done via a matrix and participants were allowed to give the same score to multiple occupations as well as decline to provide a score. The order of the occupations was randomized for each participant to avoid order fixed effects.

\VerticalSpace
We include two control occupations to ensure the quality of the responses, and both controls are listed randomly as rows among the occupations. The first control reads \textit{\textquote{Select as an answer 'four' for this question}} while the second reads \textit{\textquote{Professional athlete (e.g. competing in NFL, NBA, or NHL)}}. We discard all participants who fail to assign the rating of four to the first control, and similarly, we discard all participants who fail to assign a score less than or equal to four to the second control e.g who fail to score professional athletes as not an age-friendly occupation. We obtain 2627 valid rankings from the 210 survey participants, meaning that each of the randomly chosen occupations has been rated on average more than 87 times.

\subsection{Internal validation through regression}\label{App:Regression}

\begin{table}[!t]
\caption{AFI and Share of Older Workers in Occupations}
	\begin{adjustbox}{max totalsize = {\textwidth}{0.6\textheight}, center}
 \begin{tabular}{llllll}
\toprule
{} &       Regression \#1 &        Regression \#2 &        Regression \#3 &        Regression \#4 &        Regression \#5 \\
\midrule
AFI                                     &           $0.073^{}$ &           $0.117^{*}$ &           $0.126^{*}$ &             $0.11^{}$ &          $0.175^{**}$ \\
                                        &  $\left(0.06\right)$ &  $\left(0.064\right)$ &  $\left(0.065\right)$ &  $\left(0.067\right)$ &   $\left(0.07\right)$ \\
Share of females                        &                      &        $-0.032^{***}$ &        $-0.031^{***}$ &           $-0.016^{}$ &          $-0.038^{*}$ \\
                                        &                      &  $\left(0.011\right)$ &  $\left(0.011\right)$ &  $\left(0.016\right)$ &  $\left(0.021\right)$ \\
Share of graduates                      &                      &                       &            $-0.01^{}$ &          $-0.037^{*}$ &         $-0.045^{**}$ \\
                                        &                      &                       &   $\left(0.01\right)$ &  $\left(0.021\right)$ &  $\left(0.022\right)$ \\
Hourly wage                             &                      &                       &                       &            $0.001^{}$ &            $0.001^{}$ \\
\bottomrule
\end{tabular}
 \end{adjustbox}\VerticalSpaceFloat
\label{Tbl:regression_results_of_theta_on_maestas_index_year1990_age50_64}
{\footnotesize
    \textit{Notes:} This table shows results from regressing the 1990 share of employment of workers aged 50-64 in each occupation on the variables listed in column 1. Superscripts ***, **, and * indicate statistical significance based on a (two-sided) $t$-test using heteroskedasticity-robust standard errors at significance levels 1\%, 5\%, and 10\%, respectively.
\par}
\end{table}

\begin{table}[!t]
\caption{AFI and Share of Older Workers in Occupations}
	\begin{adjustbox}{max totalsize = {\textwidth}{0.6\textheight}, center}  \begin{tabular}{llllll}
\toprule
{} &        Regression \#1 &        Regression \#2 &        Regression \#3 &        Regression \#4 &        Regression \#5 \\
\midrule
AFI                                     &         $0.069^{***}$ &          $0.062^{**}$ &          $0.056^{**}$ &          $0.068^{**}$ &             $0.05^{}$ \\
                                        &  $\left(0.023\right)$ &  $\left(0.026\right)$ &  $\left(0.026\right)$ &  $\left(0.027\right)$ &  $\left(0.031\right)$ \\
Share of females                        &                       &            $0.005^{}$ &            $0.005^{}$ &           $-0.006^{}$ &           $-0.015^{}$ \\
                                        &                       &  $\left(0.006\right)$ &  $\left(0.006\right)$ &  $\left(0.008\right)$ &   $\left(0.01\right)$ \\
Share of graduates                      &                       &                       &            $0.006^{}$ &         $0.026^{***}$ &            $0.006^{}$ \\
                                        &                       &                       &  $\left(0.005\right)$ &  $\left(0.009\right)$ &   $\left(0.01\right)$ \\
Hourly wage                             &                       &                       &                       &         $-0.001^{**}$ &             $-0.0^{}$ \\
\bottomrule
\end{tabular}
	\end{adjustbox}\VerticalSpaceFloat
\label{Tbl:regression_results_of_theta_on_maestas_index_year1990_age65_74}
{\footnotesize
\textit{Notes:} This table shows results from regressing the 1990 share of employment of workers aged 65-74 in each occupation on the variables listed in column 1. Superscripts ***, **, and * indicate statistical significance based on a (two-sided) $t$-test using heteroskedasticity-robust standard errors at significance levels 1\%, 5\%, and 10\%, respectively.
\par}
\end{table}

\end{document}